\documentclass[11pt]{article}

\usepackage{amsmath, amssymb,amsbsy,amstext, amsthm, simplewick, amsfonts}
\usepackage{hyperref}
\usepackage{upgreek}
\usepackage{framed}
\usepackage{bbm}
\usepackage{textcomp}
\usepackage{adjustbox}
\usepackage{makecell}
\usepackage{tcolorbox}
\usepackage{physics}
\usepackage{empheq}
\usepackage[normalem]{ulem}
\usepackage{cite} 

\usepackage{braket}
\usepackage{tikz}
\usepackage{authblk}
\numberwithin{equation}{section}
\usetikzlibrary{snakes}

\newcommand{\ex}[1]{\langle #1 \rangle}
\newcommand{\be}{\begin{eqnarray} }
\newcommand{\ee}{\end{eqnarray} }
\newcommand{\bs}{\begin{split} }
\newcommand{\es}{\end{split} }

\renewcommand{\v}[1]{\mathbf{#1} }

\renewcommand{\L}{\mathcal{L}}

\newcommand{\e}{\epsilon}

\renewcommand{\O}{\mathcal{O}}

\newcommand{\psint}{\psi_{\text{int}}}
\newcommand{\psifin}{\psi}
\newcommand{\psiflat}{\psi^{\text{flat}}}
\newcommand{\dimop}{\text{dim}}
\newcommand{\Drr}{\Delta_{\text{rr}}}

\begin{document}

\begin{titlepage}
\setcounter{page}{1} \baselineskip=15.5pt 
\thispagestyle{empty}

\begin{center}
{\fontsize
{28}{28}\centering \bf A Differential Representation \\ \vspace{0.3cm} of Cosmological Wavefunctions}\\
\end{center}

\vskip 38pt
\begin{center}
\noindent
{\fontsize{12}{18}\selectfont Aaron Hillman\footnote{\tt aaronjh@princeton.edu}$^{,\star}$ and Enrico Pajer\footnote{\tt enrico.pajer@gmail.com}$^{,\dagger}$}
\end{center}

\begin{center}
\vskip 8pt
$\star$\textit{Department of Physics, Princeton University, Washington Road, Princeton, NJ, USA} \\ 
$\dagger$\textit{ Department of Applied Mathematics and Theoretical Physics, University of Cambridge, Wilberforce Road, Cambridge, CB3 0WA, UK}
\end{center}


\vspace{2.4cm}

\noindent Our understanding of quantum field theory rests largely on explicit and controlled calculations in perturbation theory. Because of this, much recent effort has been devoted to improve our grasp of perturbative techniques on cosmological spacetimes. While scattering amplitudes in flat space at tree level are obtained from simple algebraic operations, things are harder for cosmological observables. Indeed, computing cosmological correlation functions or the associated wavefunction coefficients requires evaluating a growing number of nested time integrals already at tree level, which is computationally challenging. Here, we present a new ``differential'' representation of the cosmological wavefunction in de Sitter spacetime that obviates this problem for a large class of phenomenologically relevant theories. Given any tree-level Feynman-Witten diagram, we give simple algebraic rules to write down a seed function and a differential operator that transforms it into the desired wavefunction coefficient for any scale-invariant, parity-invariant theory of massless scalars and gravitons with general boost-breaking interactions. In particular, this applies to large classes of phenomenologically relevant theories such as those described by the effective field theory of inflation or solid inflation. Trading nested bulk time integrals for derivatives on boundary kinematical data provides a great computational advantage, especially for processes involving many vertices.


\end{titlepage} 


\newpage
\setcounter{tocdepth}{2}
{
\hypersetup{linkcolor=black}
\tableofcontents
}

\newpage



\section{Introduction}


At the heart of our current cosmological paradigm sits the observation that the distributions of everything we observe on cosmological scales, from galaxies to dark matter, from neutrinos to photons, was seeded by quantum fluctuations during a primordial phase that preceded the hot big bang. To make predictions for cosmological surveys requires computing the statistics of these fluctuations using quantum field theory in curved spacetime and accounting for the quantum behavior of  spacetime (metric) fluctuation. A main driver for this line of research is the expectation that current or future cosmological datasets will shed light on new physics beyond the standard model and perhaps the perturbative regime of quantum gravity.\\

The workhorse of this program is perturbation theory, which has been used extensively in the past two decades to compute predictions for a variety of models and classes of theories. On the one hand, perturbation theory gives us with a very rich set of explicit and phenomenologically useful results and predictions. On the other hand, it is also a springboard to the exploration of deeper non-perturbative structures. For example, much of our understanding of non-perturbative results in quantum field theory rests on extrapolating the observed behavior in perturbation theory. One example are positivity bounds \cite{Positivity1}, where we extrapolate the analytic structure of perturbation theory to the UV-completion of effective field theories and use it to discriminate theories that can be extended to arbitrarily high energies in a consistent way. Another example is the S-matrix numerical bootstrap approach to the calculation of non-perturbative amplitudes (see e.g. \cite{Paulos:2016fap}). These successes in the study of flat-space observables motivated a recent resurgence in the study of \textit{perturbation theory} on de Sitter and cosmological spacetimes \cite{cosmopoly,Goon:2018fyu,Arkani-Hamed:2018bjr,Benincasa:2018ssx,CosmoBootstrap1,Baumann:2019oyu,Sleight:2019mgd,Sleight:2019hfp,Benincasa:2019vqr,Bzowski:2019kwd,Baumann:2020dch,Green:2020ebl,COT,Sleight:2020obc,BBBB,MLT,Isono:2020qew,Melville:2021lst,Goodhew:2021oqg,Bonifacio:2021azc,DiPietro:2021sjt,Sleight:2021plv,Hogervorst:2021uvp,Meltzer:2021zin,Sleight:2021iix,Cespedes:2020xqq,Gomez:2021qfd,Cabass:2021fnw} which builds upon many results obtained over the past twenty years \cite{Maldacena:2002vr,WFCtoCorrelators1,Mata:2012bx,Bzowski:2013sza,Kundu:2014gxa,Kundu:2015xta}. One of the eventual goals of this program is deriving cosmological positivity bounds\footnote{Several constraints have been derived in flat space for boost-breaking effective theories that capture in the sub-Hubble limit of common models of inflation and dark energy, see e.g. \cite{Conjecture,TanguyScott,PSS,DSJS,Positivity2,Melville:2019wyy,Stefanyszyn:2020kay}.} and a non-perturbative cosmological bootstrap (see \cite{Celoria:2021vjw,DiPietro:2021sjt,Hogervorst:2021uvp} for progress in this direction).\\

In this work, we derive a new representation of tree-level wavefunction coefficients for massless scalars and gravitons to any order in perturbation theory. We call this a ``differential representation'' because the wavefunction coefficients are obtained by acting with  differential operators on a class of simple seed functions, which in turn are determined by purely algebraic rules. Our differential representation covers all boost-breaking interactions that appear in phenomenological models of inflation and dark energy, as for example in the effective field theory of inflation \cite{Creminelli:2006xe,Cheung:2007st,Bordin:2018pca} or solid inflation \cite{SolidInflation}. Following the hint of cosmological observations, we assume scale invariance throughout our work, but we neglect the small deviation implied by the measured scalar spectral tilt. In practice, given a certain Feynman-Witten diagram in dS, where interactions are specified for each vertex, our differential representation generates a seed function and a differential operator that acts on it to give the associated wavefunction coefficient. All differential operators act on the external spatial momenta of the Fourier-space wavefunction coefficient and consequently the notion of time in the bulk of de Sitter completely disappears from the calculation. This representation turns out to be much simpler than the lengthy calculation of nested time integrals and even much more compact that the result of those integrals. The consequent improvement in calculation time is already remarkable for a single exchange diagram and grows exponentially with the number of internal legs. For example, we show that the differential representation of a five-point function with three cubic $  \phi'^{3} $ couplings and two internal lines (three-site chain) is contained in less than a line. For the reader familiar with our results, this can be computed with pen and paper in minutes, while performing the corresponding time integrals in the standard bulk-representation requires much longer with an off-the-shelf software such as Mathematica on a standard CPU. \\

\begin{figure}
	\begin{center}
		\begin{tikzpicture}
		    \coordinate (c1) at (0, 0);
			\draw (c1) circle (1 cm);
			\node at (c1) {\LARGE $\psiflat$};
			\coordinate (a1L) at (1.4, 0);
			\coordinate (a1R) at  (3, 0);
			\draw[thick, ->] (a1L) -- (a1R);
			\coordinate (c2) at (4.5, 0);
			\draw (c2) circle (1 cm);
			\node at (c2) {\LARGE $ \psint$};
			\coordinate (t1) at (0, -1);
			\coordinate (t2) at (0, -2);
			\coordinate (b1) at (-1, -2);
			\coordinate (b2) at (1, -2);
			\coordinate (b3) at (-1, -2.5);
			\coordinate (b4) at (1, -2.5);
			\node at (0, -2.25){$ \begin{array}{c} \text{Flat space seed} \\ \text{(Sec.\ref{sec:3})}\end{array} $};
			\coordinate (t3) at (4.5, -1);
			\coordinate (t4) at (4.5, -2);
			\coordinate (b5) at (3.5, -2);
			\coordinate (b6) at (5.5, -2);
			\coordinate (b7) at (3.5, -3);
			\coordinate (b8) at (5.5, -3);
			\node at (4.5, -2.3) {$ \begin{array}{c} \text{Collapsing \&} \\ \text{re-routing (Sec.\ref{sec:4})} \end{array}$  };
			\coordinate (a2L) at (5.9, 0);
			\coordinate (a2R) at  (7.5, 0);
			\draw[thick, ->] (a2L) -- (a2R);
			\coordinate (c3) at (9, 0);
			\draw (c3) circle (1 cm);
			\node at (c3) {\LARGE $ \psifin$};
			\coordinate (b9) at (7, -2);
			\coordinate (b10) at (11, -2);
			\coordinate (b11) at (11, -3.6);
			\coordinate (b12) at (7, -3.6);
			\coordinate (t5) at (9, -1);
			\coordinate (t6) at (9, -2);
			\node at (9, -2.3) { $ \begin{array}{c} \text{External lines} \\ \text{\& vertices (Sec.\ref{sec:5})} \end{array}  $};
			\node at (2.2, .4) {$\Drr$};
			\node at (6.7, .4) {$ F \Delta_A$};
			\node at (0, 1.6) {Step 1};
			\node at (4.5, 1.6) {Step 2};
			\node at (9, 1.6) {Step 3};
		\end{tikzpicture}
	\end{center}
	\caption{Depiction of the sequence of differential operators acting on the flat space seed function $\psiflat$ to obtain the function $\psifin$ specified by the edges and vertices.\label{fig1}}
\end{figure}
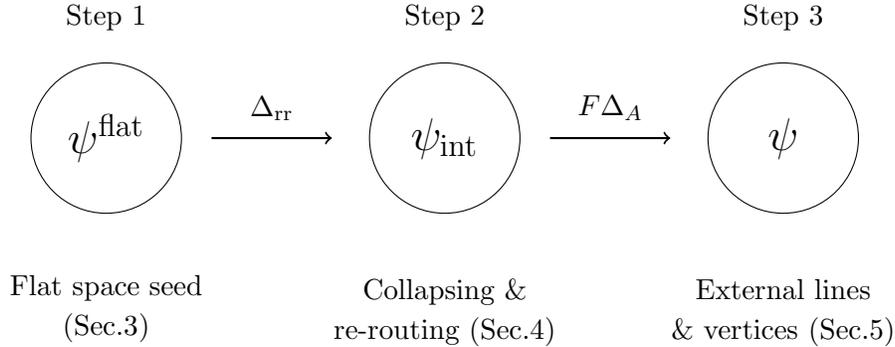

The rest of the paper is organized as follows. For the reader interested in using our results in practical calculations, in Section \ref{sec:2} we give a brief summary of our three-step procedure to derive the differential representation of any tree-level contribution to the $  n $-th wavefunction coefficient  $  \psi_{n} $ for a massless scalar field in de Sitter. The following three sections provide a derivation of each of the steps of this procedure, as depicted in Figure \ref{fig1}, and a number of practical examples. In particular, in Section \ref{sec:3}, we introduce the seed wavefunction $  \psiflat $, which is the wavefunction of a massless scalar with polynomial interactions in flat spacetime (Step 1). Then, in Section \ref{sec:4}, we derive the differential operator that acting on $  \psiflat $ gives us an intermediate wavefunction $  \psint $, which accounts for the difference in mode functions between flat spacetime and de Sitter (Step 2).  At the level of the bulk time integral, this step correctly reproduces the integrand modulo overall powers of conformal time at each vertex.  The last step is discussed in Section \ref{sec:5} and consists of fixing the correct overall powers of conformal time at each vertex as well as adding the appropriate momentum factors to the vertices and the external legs, yielding the final de Sitter wavefunction $  \psifin $ (Step 3). In Section \ref{sec:minimal} we discuss the extension of our procedure to include massless gravitons and the interactions of minimal coupling to gravity with two explicit examples. We conclude in Section \ref{sec:end} with a discussion and an outlook. Appendix \ref{app:A} contains the computational details of relations obeyed by propagators that we use in the bulk of the paper. \\

While this paper was (slowly) being completed, reference \cite{Baumann:2021fxj} appeared, which also develops the idea of relating the dS wavefunction to flat spacetime objects with differential operators.

\paragraph{Notation and conventions} We will use the following de Sitter line element (in mostly positive signature)
\begin{align}
ds^{2}=-dt^{2}+a^{2}dx^{i}\delta_{ij}dx^{j}=\frac{-d\eta^{2}+dx^{i}\delta_{ij}dx^{j}}{H^{2}\eta^{2}}\,,
\end{align}
with scale factor $  a=e^{Ht}=-1/(H\eta) $. 

Our bulk-to-bulk propagator $  G $ has an overall factor of $  i $, schematically $  G\sim i P K K $ (see \eqref{flatprop}). With this convention the Feynman rules are that every internal line corresponds to a $  G $, there is an overall $ i $ for every diagram, and vertices do not get any $  i $, for example the vertex of $g\phi^{3}/3! $ is simply $g  $. 

We use $  \v{k} $ and $  k=|\v{k}| $ for \textit{external momenta} labelled by letters in the first half of the latin alphabet, $  a,b=1,\dots,n $ for an $  n $-point function. The $  I $ \textit{internal momenta} are denoted by $  \v{p}_{m} $ and their energy is $  p_{m} = |\v{p}_{m}|$ with $  m=1,\dots,I $. Finally, the $  V=I+1 $ vertices have valency $  n_{A} \geq 3 $ (the number of external and internal legs ending on the vertex) and are placed at conformal time $  \eta_{A} $ with $  A=1,\dots,V $. The sum of energies $k_a$ running from a single vertex $A$ to the boundary is denoted by $q_A = \sum\limits_{a \in A}k_a$.

The wavefunction for some future boundary configuration of a field $\Phi(\vec x)$ is given by a path integral and is naturally organized as
\begin{equation}
	\Psi[\Phi] = \exp\left\{-\sum\limits_n \, \frac{1}{n!} \int \left[ \prod_i^{n}\frac{ d^3k_i}{(2\pi)^{3}} \Phi(\v{k}_i) \right] \psi_n\,(2\pi)^{3}\delta^{3}\left(  \sum \v{k}_{a}\right)\right\}\,.
\end{equation}
The delta-function stripped $\psi_n$'s are the wavefunction coefficients we are interested in computing and can be defined as 
\begin{equation}
	(2\pi)^{3}\delta^{(3)}\bigg(\small\sum\normalsize\limits_i \vec k_i \bigg)\psi_n = \frac{\delta^n \Psi}{\delta \Phi(\v{k}_1)\dots \delta \Phi(\v{k}_n)}\bigg |_{\Phi = 0} \,.
\end{equation}
These can be computed in perturbation theory using Feynman-Witten rules. We will indicate by $  \psi_{n} $ the wavefunction coefficients in de Sitter, while $  \psiflat_{n} $ is reserved for the wavefunction coefficients in flat spacetime of a theory with only polynomial interactions (see Sec. \ref{sec:3}). The kinematic dependence of $  \psiflat $ can be paramerized in terms of the combinations $  q_{A} $ for each vertex $  A $ and the edge energies $p_m$. Therefore, instead of writing $  \psiflat(k_{1},\dots,k_{n}) $, we will use the notation $  \psiflat(q_{1},\dots,q_{V}; p_1\dots, p_I) $, where $  V $ is the number of vertices and $I$ the number of edges. In general $  \psiflat $ depends on the internal energies $  p $, but we will always omit this dependence when no ambiguity arises.

\section{Overview}\label{sec:2}

In this section, we give an overview of our procedure to compute the cosmological wavefunction for any given tree-level diagram for a massless scalar field in de Sitter with arbitrary boost-breaking local interactions that respect scale invariance. In particular our approach does not rely on invariance under de Sitter boosts and can be applied for example to any of the interactions resulting from the Effective Field Theory of Inflation \cite{Creminelli:2006xe,Cheung:2007st}. Our procedure consists of three steps. In Step 1 we start from a seed wavefunction $  \psiflat $ that can be computed \textit{algebraically} using the results of \cite{cosmopoly}.  These functions are determined by the topology of the amputated diagram i.e. the diagramed defined by only the internal lines (edges).  Depending on the details of the interactions, additional $  \psiflat $'s diagrams are needed corresponding to different ways of collapsing internal lines.  In Step 2 we give a prescription to write down a derivative operator that acts on $  \psiflat $ and accounts for the structure of the exchange propagators up to overall powers of conformal time in the integrand. Finally, in Step 3 we fix the correct power of conformal time at each vertex and account for external kinematical factors coming from spatial derivatives and polarization tensors.

In its first and most basic implementation, our approach works only for interactions with enough derivatives to cancel all inverse powers of $  \eta $ present in the measure of covariant spacetime integrals. More precisely, we start by considering vertices that satisfy the following criterion
\begin{equation}
\label{bound}
	d\equiv 2 n_{\partial_t}+n_{\partial_i} \geq 4\,,
\end{equation} 
where $  n_{\partial_t} $ and $ n_{\partial_i}  $ are the number of time and spatial derivatives, respectively. For scalar perturbations this is not a very strong restriction. For example, we know that the largest interactions in a generic model of single-field inflation can be captured by the decoupling limit of the EFT of inflation, where a shift symmetry is imposed on the Goldstone boson of time translations. The leading interactions relevant for the bispectrum are $  \dot \phi^{3} $ and $  \dot \phi \partial_{i}\phi^{2} $, which in turn lead to equilateral and orthogonal non-Gaussianities \cite{Senatore:2009gt}. Both of these interactions satisfy the above criterion and so do all the shift symmetric operators contributing to quartic and higher interactions. Later, in Section \ref{sec:minimal}, we will show that the restriction in \eqref{bound} can be relaxed in certain cases, such as the minimal coupling to gravity. In the following we summarize the three steps involved in deriving the differential representation of the dS wavefunction.\\


\paragraph{Step 1: The seed functions from flat spacetime} The first step consists in computing the wavefunction coefficient in Minkowski for a theory with purely polynomial interactions. A simple and purely algebraic prescription to write down this ``seed'' function $  \psiflat $ without explicitly performing any time integral was described in \cite{cosmopoly} and we review it in detail in Section \ref{sec:3}. Any tree-level wavefunction coefficient $  \psiflat_{V} $ is represented by a tree diagram with $  V $ vertices and $ I=V-1   $ internal lines (edges). To every vertex $  A=1,\dots,V $ we associate a total external energy $  q_{A} $, while to every edge $  m=1,\dots,I $ we associate an internal energy $  p_{m} $. The algebraic computation of $  \psiflat $ is based on the following recursive relation
\begin{align}
\left(\sum\limits_A^{V} q_A \right)\psiflat_{V}=\sum_{m}^{I}\psiflat_{V'}(q_{1},\dots,q_{B}+p_{m},\dots,q_{n'})\psiflat_{V-V'}(q_{1},\dots,q_{B'}+p_{m},\dots,q_{n-n'})
\end{align}
where the sum runs over all edges $  m=1,\dots,I $ and $ 0<V'<V$ is the number of vertices left in one of the two subdiagrams when the $ m$-th edge connecting vertices $B $ and $ B'$ is removed (see \eqref{graphicalrep} for a graphical representation)\footnote{$V$ of course does not uniquely define a topology for $V > 3$ and it is implied that the we have the graphs induced by the original graph when cut on edge $m$.  This is made clear in \eqref{graphicalrep}.}.  Our analysis is restricted to tree level and so we have omitted the loop-deletion term in the above recursion. From this relation we can easily write down all desired $  \psiflat $ starting from the trivial $  V=1 $ base case. For example\footnote{Notice that where no confusion arises we indicate explicitly only the dependence of $  \psiflat $ on the external energies $  q_{A} $ and leave the dependence on the internal energies $  p_{m} $ implicit.}:
\begin{align}
\psiflat_{1}(q)&=  \frac{1}{q}\,, \\
\psiflat_{2}(q_{1},q_{2};p) &= \frac{\psiflat_{1}(q_{1}+p)\psiflat_{1}(q_{2}+p)}{q_1+q_2}\,,\\
\psiflat_{3}(q_{1},q_{2},q_{3};p_{1},p_{2})&=\frac{ \psiflat_{2}(q_{1},q_{2}+p_{2})\psiflat_{1}(q_{3}+p_{2})+\psiflat_{1}(q_{1}+p_{1})\psiflat_{2}(q_{2}+p_{1},q_{3})}{ \sum_{A}^{3} q_{A}}\,,
\end{align}
and so on and so forth.

\begin{figure}
	\begin{center}
		\begin{tikzpicture}[scale=1.25]
			\draw[thick] (0, 0) circle (5.5mm);
			\draw[thick] (4, 0) circle (5.5mm);
			\draw[thick] (.55, 0) -- (3.45, 0);
			\node at (3, -.3) {$\phi$};
			\draw[->] (2.7, .4) -- (3.3, .4);
			\draw [->] (1.3, .4) -- (.7, .4);
			\draw[dashed] (2.4, 1.4) -- (2.4, -1);
			\node at (.7, 1.1) {\small$\text{opposite-vertex side}$};
			\node at (3.2, 1.1) {\small$\text{vertex side}$};
			\node at (0, 0) {$G_L$};
			\node at (4, 0) {$G_R$};
		\end{tikzpicture}
	\end{center}
	\caption{For a half-edge $\phi$ the graphic defines flowing vertex side or opposite-vertex side.  A choice of flow determines a differential operator summing over energies in $G_L$ including the edge itself or in $G_R$ including the vertex.\label{routing}}
\end{figure}
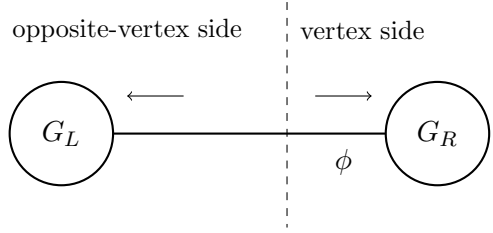

\paragraph{Step 2: Intermediate Wavefunction} In Step 2, a new ``intermediate'' wavefunction $\psint$ is obtained from $\psiflat$ by applying a suitable sequence of differential operators that are fully fixed by specifying where time derivatives at each vertex act on an internal bulk-to-bulk propagator. Since each internal line or ``edge'' is attached to two vertices, it is useful to speak in terms of ``half-edges'' (see Figure \ref{routing}) which are uniquely specified by a vertex edge pairing.
We may therefore denote a half-edge  by a pair of indices $ mA$, where the first index indicates that the half-edge is part of the $ m$ edge and the second index indicates that it ends on the vertex $ A$. An half-edge that has no time derivatives is said to be of the $ \phi$ type, while an half-edge with one time derivative is said to be of the $ \phi'$ type. As we are considering interaction vertices with at most one time derivative\footnote{In perturbation theory, any higher time derivative can always be re-written in terms of at most one time derivative using repeatedly the (non-linear) equations of motion. This procedure also generates contributions to the wavefunction from field redefinitions (we thank S. Jazayeri for pointing this out), which don't have total energy poles and correspond to contact terms in position space. These contributions from field redefinitions are neglected in our analysis. If desired, they can be included straightforwardly with traditional methods.} there are no other possibilities. Once the two half-edges of a given edge are specified, four options for the corresponding bulk-to-bulk propagators $  G(\eta,\eta',p) $ emerge: no time derivatives, which we indicate by $\braket{\phi \phi}$, one time derivative,   $ \braket{\phi' \phi}  $ or $ \braket{\phi \phi'}  $, or two time derivatives, $\braket{ \phi' \phi'}$. With these definitions we have\footnote{Here $ \eta'$ is just another name for a time variable. The prime should not be confused with the prime on $ \phi$, which indicates a time derivative, $ \phi'=\partial_{\eta}\phi$.}
\begin{align}\label{props}
&\begin{tikzpicture}[baseline=(current  bounding  box.center)]
		\coordinate (v1) at (-6, 0);
		\coordinate (v2) at (-2, 0);
		\draw[thick] (v1) -- (v2);
		\draw[fill] (v1) circle (.5mm);
		\draw[fill] (v2) circle (.5mm);
		\node at (-5.75, .3) {$ \phi$};
		\node at (-2.25, .3) {$ \phi$};
		\node at (3, 0) {$ G_{\phi\phi}(\eta, \eta', p )= G(\eta, \eta', p)  \,,  $};
		\node at (-6, -.3) {$\eta$};
		\node at (-2, -.3) {$\eta'$};
		\node at (-4, -.3) {$p$};
	\end{tikzpicture} \\
&\begin{tikzpicture}[baseline=(current  bounding  box.center)]
		\coordinate (v1) at (-6, 0);
		\coordinate (v2) at (-2, 0);
		\draw[thick] (v1) -- (v2);
		\draw[fill] (v1) circle (.5mm);
		\draw[fill] (v2) circle (.5mm);
		\node at (-5.75, .3) {$ \phi$};
		\node at (-2.25, .3) {$ \phi'$};
		\node at (3, 0) {$ G_{\phi \phi'}(\eta, \eta', p) = \partial_{\eta'} G(\eta, \eta', p) \,, $};
		\node at (-6, -.3) {$\eta$};
		\node at (-2, -.3) {$\eta'$};
		\node at (-4, -.3) {$p$};
	\end{tikzpicture} \\
&\begin{tikzpicture}[baseline=(current  bounding  box.center)]
		\coordinate (v1) at (-6, 0);
		\coordinate (v2) at (-2, 0);
		\draw[thick] (v1) -- (v2);
		\draw[fill] (v1) circle (.5mm);
		\draw[fill] (v2) circle (.5mm);
		\node at (-5.75, .3) {$ \phi'$};
		\node at (-2.25, .3) {$ \phi'$};
		\node at (3, 0) {$ G_{\phi'\phi'}(\eta, \eta', p) = \partial_{\eta}\partial_{\eta'} G(\eta, \eta', p)\,.  $};
		\node at (-6, -.3) {$\eta$};
		\node at (-2, -.3) {$\eta'$};
		\node at (-4, -.3) {$p$};
	\end{tikzpicture} 
\end{align}
Given the explicit form of $  G $ for fields with the massless de Sitter mode function in \eqref{G}, these propagators obey the following relations to the flat space propagator  
\begin{align}\label{props}
G_{\phi\phi}(\eta, \eta', p) &= \frac{1}{p^2}\left[ (1-\eta\partial_\eta)(1-\eta'\partial_{\eta'})G_{\text{flat}}(\eta, \eta', p)+\eta\eta'\delta(\eta-\eta')\right]\,,\\
G_{\phi'\phi'}(\eta, \eta', p) &= \eta\eta'\left[p^2G_{\text{flat}}(\eta, \eta', p)-\delta(\eta-\eta') \right]\,,\label{propspp} \\
G_{\phi \phi'}(\eta, \eta', p) &= (1-\eta\partial_\eta)\eta' G_{\text{flat}}(\eta, \eta', p) =G_{\phi' \phi}(\eta', \eta, p) \,.\label{propsp}
\end{align}
It is straighforward to keep track of the powers of energy $ p$. This requires multiplying by the factor specified in \eqref{pfactor}. On the other hand, there are two aspects of the above relations that require more work. We have to account for (\textit{i}) the relative powers of $\eta$ in the terms of the exchange propagators and (\textit{ii}) for the $\delta$ functions. Issue (\textit{i}) will be addressed by the operation of \textit{re-routing}. This is a manifestation of integration by parts in terms of a differential operator which captures how the time derivatives get routed out to a boundary vertex. Issue (\textit{ii}) will be addressed by the operation of \textit{collapsing}. Each term in the sum over ways of collapsing edges will get its own rerouting operator. \\

Let's start with re-routing. Using integration by parts, the temporal differential operators $(1-\eta\partial_\eta)$ can be routed out of all time integrals to act on the times of those vertices of the diagram that are only connected to a single internal edge. When acting on these times, the derivatives can be re-expressed as kinematic differential operators on $\psiflat$.  In \eqref{props}-\eqref{propsp} we see that a $ \partial_{\eta}$ appears for every half-edge $\phi$ in the diagram. Moreover, as discussed below, we must also sum over diagrams with $\braket{ \phi' \phi'}$ and $\braket{\phi \phi}$ propagators collapsed by the $\delta$-functions. The differential operator $ \Drr$ (``rr'' stands for re-routing) which accounts for re-routing $(1-\eta\partial_\eta)$ on a given $\phi$ half-edge is (see Figure \ref{routing})
\begin{align}\label{rerouting}
	\begin{cases}\Drr = 1+\left[\sum\limits_A \left(1+q_A\partial_{q_A}\right)+\sum\limits_m \left( 1+ p_m\partial_{p_m} \right) \right] & \text{vertex side}\,, \\\\ \Drr = 1-\left[\sum\limits_A \left(1+q_A\partial_{q_A}\right)+\sum\limits_m \left( 1+p_m\partial_{p_m} \right)\right] &\text{opposite-vertex side}\,,\end{cases} \hspace{3mm}
\end{align}
where the sum runs over all vertices encountered when flowing out of the diagram from the chosen half edge. In general, one needs a $ \Delta_{\text{rr}}$ operator for each $ \phi$ half-edge in the diagram and when necessary we will specify which half-edge $ \Delta_{\text{rr}}$ corresponds to, as e.g. in \eqref{spec1}, \eqref{spec2} and \eqref{deltagrav}. The order of application of these differential operators is relevant and can be understood from the integration by parts.  Specifically, on a given path out of the diagram, the operator corresponding to the innermost $\phi$ half-edge should be applied to the seed function $\psiflat$ first. Moreover, when multiple $\phi$ half-edges are present, choices of flow should not cross. This will be expanded upon in Section \ref{sec:4}. \\

Let's move on to collapsing. Two of the three propagator relations have $\delta$-functions: $G_{\phi\phi}$ and $G_{\phi' \phi'}$. Each $\delta$-functions has the effect of collapsing the corresponding edge and merging the two vertices to which that edge was attached. For example a $ \delta(\eta_{A}-\eta_{B})$ would merge vertex $ A$ at time $ \eta_{A} $ with external energy $ q_{A}$ with vertex $ B$ at time $ \eta_{B}$ with external energy $ q_{B}$. In doing this we have to be careful with the overall powers of conformal time. The $\delta$-functions in $G_{\phi\phi}$ and $G_{\phi' \phi'}$ have different net powers of conformal time relatively to the non-$\delta$-function terms in each expression. To account for this difference, we count the powers of conformal time in the $\delta$-function terms relatively to the flat-space propagator. For $\phi'\phi'$, the power is the same and therefore the collapsed contribution has merely a minus sign. For $\phi\phi$, the $ \delta$-function has two powers of conformal time in excess of term with $ G_{\text{flat}}$ and therefore collapsing the edge should be accompanied by a single derivative with respect to each of the vertex energies to which the edge is connected, namely $ (-i\partial_{q_{A}})(-i\partial_{q_{B}})$, or equivalently two derivatives of the energy of the merged vertex. These operations act after the re-routing operators defined below. \\

\indent Lastly, we account for the power spectrum normalization of a given (in general collapsed) diagram.  This factor is merely the product over $p_m^2$ for each (non-collapsed) $\phi'\phi'$ exchange edge and $p_m^{-2}$ for each (non-collapsed) $\phi\phi$ exchange edge.  Equivalently, this is the product over $p_m$ for each $\phi'$ in the truncated\footnote{All external lines should be amputated here because they are captured later in Step 3.} diagram and $p_m^{-1}$ for each $\phi$.  We can call this
\begin{equation}
\label{pfactor}
\text{P} = 	\prod\limits_{\phi'\phi'}p_m^2 \prod\limits_{\phi\phi}\frac{1}{p_m^2}
\end{equation}
where the products are over propagators of each type.\\ 

Summarizing, in Step 2 each internal propagator requires the following operations
\begin{align}
\phi' \phi' &\to \text{collapsing}\,,&
\phi \phi' &\to \text{re-routing}\,,&
\phi \phi &\to \text{collapsing \& re-routing} \,.
\end{align}
For every collapsed edge with energy $ p$ that merges vertex $ A$ and $ B$, the following operator must be applied after the application of re-routing operators:
\begin{align}
\text{collapsing } \phi\phi &\to	-\frac{1}{p^2}\partial_{q_{A}}\partial_{q_{B}}  & \text{collapsing } \phi'\phi' &\to	-1 \,.
\end{align}
Notice that the re-routing operation is not required for the collapsed contribution, it only appears in the non-collapsed contribution. The intermediate wavefunction is comprised of the sum over all ways of collapsing the relevant propagators and applying the prescribed differential operators on the seed wavefunction in each case.  Each diagram in the sum over collapsing should be multiplied by the associated P-factor \eqref{pfactor}.

\paragraph{Step 3: Final de Sitter Wavefunction  } In the final step we derive the desired de Sitter wavefunction coefficient. We now specify the precise structure of the vertices and bulk-to-boundary propagators, not merely the skeleton and structure of the internal lines (a.k.a. bulk-to-bulk propagators).  To this end, we apply derivatives with respect to the vertex energies appearing in the bulk-to-boundary propagators, we add factors of $  k_{a}^{2} $ or $(1-k_i \partial_{k_i})$ to reproduce the external massless bulk-to-boundary mode functions with one time-derivative ($ K'  $) or no time-derivatives ($  K $), respectively. In particular, for every vertex we apply the following operator (see Sec \ref{sec:5})
\begin{equation}
		\Delta_A = i^{4-d_A}\left( \prod_{a\in K'} k_{a}^{2} \right) \left( \prod_{b \in K} \left(1-k_b \partial_{q_A}\right) \right) \frac{\partial^{d_A-4}}{\partial q_A^{d_A-4}}\,,
\end{equation}
where $  d_{A} $ was defined in \eqref{bound}. 

We also account for all contractions of spatial momenta and polarization tensors by adding a multiplicative factor $  F $. In summary, the final de Sitter wavefunction coefficient is given by
\begin{align}
	\psifin = F \left( \prod\limits_A \Delta_A \right)   \psint\,.
\end{align}


\begin{figure}
	\begin{center}
		\begin{tikzpicture}
			\coordinate (b1) at (-6, 0);
			\coordinate (b2) at (0, 0);
			\coordinate (b3) at (6, 0);
			\coordinate (b4) at (-3, 0);
			\coordinate (b5) at (3, 0);
			\coordinate (v1) at (-4, -2);
			\coordinate (v2) at (0, -2);
			\coordinate (v3) at (4, -2);
			\draw[thick] (b1) -- (v1) -- (v2) -- (v3) -- (b3);
			\draw[thick] (b4) -- (v1);
			\draw[thick] (b5) -- (v3);
			\draw[thick] (b2) -- (v2);
			\draw[very thick] (-7, 0) -- (7, 0);
			\node at (b1) [anchor=south]{$  \phi(\v{k}_{1}) $};
			\node at (b3) [anchor=south]{$  \phi(\v{k}_{5}) $};
			\node at (b2) [anchor=south]{$  \phi(\v{k}_{3}) $};
			\node at (b4) [anchor=south]{$  \phi(\v{k}_{2}) $};
			\node at (b5) [anchor=south]{$  \phi(\v{k}_{4}) $};
			\node at (-4.75,-1.65) {$\dot \phi $};
			\node at (-3.5, -2.35){$ \partial_i\phi$};
			\node at (-3.4, -1.55){$ \partial_i\phi$};
			\node at (-0.45, -2.35){$ \partial_j\phi$};
			\node at (0.45, -2.35){$ \partial_j\phi$};
			\node at (0.25, -1.55){$ \dot\phi$};
			\node at (4.75,-1.65) {$\dot \phi $};
			\node at (3.5, -2.35){$ \dot\phi$};
			\node at (3.4, -1.55){$ \dot\phi$};
			\node at (-2, -1.7) {$p_1$};
			\node at (2, -1.7) {$p_2$};
		\end{tikzpicture}
	\end{center}
\caption{A particular contribution to the quintic wavefunction coefficient from the indicated cubic interactions.\label{overviewexample}}
\end{figure}
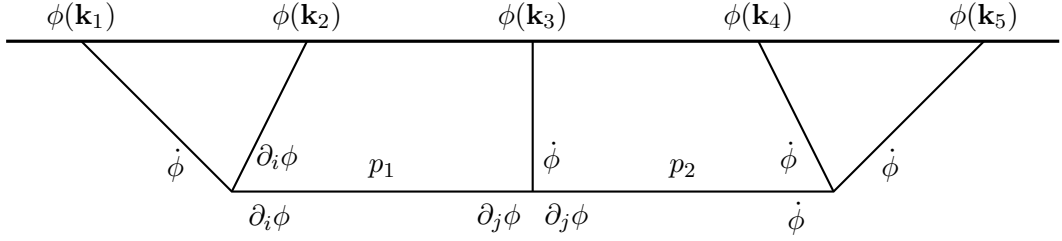


\paragraph{Featured examples} Now we present two examples to illustrate the application of the above prescription. In the first example, we will calculate the contribution coming from the diagram depicted in Fig. \ref{overviewexample}.  This does not represent the full contribution at this order in the couplings; that will be presented in Section \ref{sec:5}.  Nonetheless, this example illustrates the essential content of the prescribed rules.\\
\indent The relevant seed functions are
\begin{align}
	\psi_3^{\text{flat}}(q_1, q_2, q_3; p_1, p_2) &= \frac{1}{q_T(q_1+p_1)(q_3+p_2)(q_2+p_1+p_2)}\left(\frac{1}{q_1+q_2+p_2}+ \frac{1}{q_3+q_2+p_1}\right) \,,\\
	\psi_2^{\text{flat}}(q_1+q_2, q_3; p_2) &= \frac{1}{q_T(q_1+q_2+p_2)(q_3+p_2)}\,,
\end{align}
where $  q_{1}=k_{1}+k_{2} $, $  q_{2}=k_{3} $, $  q_{3}=k_{4}+k_{5} $ and $q_T = q_1+q_2+q_3$.  The intermediate wavefunction is then (see \eqref{introref})
\begin{multline}
\psi_{\text{int}} = \frac{1}{p_1^2}(2+q_1\partial_{q_1})(-1-p_1\partial_{p_1}-q_1\partial_{q_1})(4+q_2\partial_{q_2}+p_1\partial_{p_1}+q_1\partial_{q_1})\psi_3^\text{flat}\\
-\frac{1}{p_1^2}\partial_{q_1}\partial_{q_2}(2+q_1\partial_{q_1})\psi_2^{\text{flat}}\,.
\end{multline}
Finally the desired contribution to the dS wavefunction coefficient is
\begin{equation}
	\psi = -(\v{p}_1\cdot \v{k}_2)(\v{p}_1\cdot \v{p}_2)(k_1 k_3 k_4 k_5)^2(1-k_2\partial_{q_1})\partial_{q_3}^2\psi_{\text{int}}\,.
\end{equation}

In the second example, we provide a simple formula for \textit{any} tree-level diagram with $  n $ external legs, $  V  $ vertices and $  I=V-1 $ internal lines in a theory where all interactions are of the form $  \phi'^{n} $ for some integer $  n\geq 3 $. The corresponding contribution to the dS wavefunction coefficient $  \psi_{n} $ is given by 
\begin{align}
\psi_{n}=\prod_{a=1}^{n} k_{a}^{2}   \prod_{A=1}^{V}(-1)^{n_A}\partial_{q_{A}}^{2n_{A}-4}\, \sum_{\text{collapsings}}^{2^{I}} \left[ (-1)^{n_{c}}\psiflat_{n-n_{c}} \prod_{m}^{I-n_{c}}p_{m}^{2}\right]\,,
\end{align}
where the sum is over the $  2^{I} $ possible ways to collapse a number $0\leq n_{c} \leq n $ of internal lines and the product goes over the $  I-n_{c} $ un-collapsed internal lines. Here $  n_{A} $ is the valency of each $  \phi'^{n_{A}} $ interaction and the arguments of the seed wavefunctions $  \psiflat $ are determined by the specific collapsing under consideration.


\section{Step 1: The seed flat-space wavefunction $\psiflat$}\label{sec:3}

In this section, we discuss the seed function $\psiflat$ and review the simple algebraic prescription derived in \cite{cosmopoly} to compute it for any tree-level diagram\footnote{The results of \cite{cosmopoly} also apply to loop-level integrands, but we restrict our analysis to tree level.}.\\

Let the ``seed'' wavefunction $  \psiflat $ be the wavefunction for a massless scalar field in Minkowksi spacetime with polynomial interactions (no derivatives) at time $  t=0 $. In the next steps we derive the desired cosmological wavefunction by acting with a differential operator on $  \psiflat $. At tree-level, contributions to a wavefunction coefficient $  \psiflat_{n} $ are represented by a tree diagram with $  V $ vertices, $  I=V-1 $ edges (internal lines), and a number of external lines. The ``bulk'' representation of the corresponding wavefunction is given in terms of the following integral expression 
\begin{equation}\label{seed}
	 \psiflat(\{k_n\}; \{ p_m\}) = i\int d^V t \left[  \prod_{a=1}^{n}K(k_{a})\right] \left[ \prod_m^{I}  G_{\text{flat}}( p_m) \right]\,,
\end{equation}
where the so-called \textit{bulk-to-boundary propagator} $  K(k,\eta) $ for a massless scalar in Minkowski is written in terms of the Minkowski mode functions $  \phi^{+}_{\text{flat}} $ as
\begin{align}\label{Kflat}
K_{\text{flat}}(k,t)=\frac{\phi^{+}_{\text{flat}}(k,t)}{\phi^{+}_{\text{flat}}(k,0)}=\frac{\frac{1}{\sqrt{2k}}e^{ikt}}{\frac{1}{\sqrt{2k}}e^{ik0}}=e^{ikt}\,.
\end{align}
The so-called \textit{bulk-to-bulk propagator} $  G_{\text{flat}} $ contains a term proportional to the usual Feynman propagator but also an additional term that ensures the vanishing of $  G_{\text{flat}} $ as $t, t' \to 0$. It is explicitly given by  
\begin{align}\label{flatprop}
G_{\text{flat}}(t, t', k) &=iP_{\text{flat}}(k)\left[  \theta(t-t') K_{\text{flat}}^{\ast} \nonumber(t)K_{\text{flat}}(t')+\theta(t'-t) K_{\text{flat}}^{\ast}(t')K_{\text{flat}}(t)-K_{\text{flat}}(t)K_{\text{flat}}(t')\right] \\ \nonumber
&=i\left[  \theta(t-t') \phi_{\text{flat}}^{-}(t)\phi_{\text{flat}}^{+}(t')+\theta(t'-t) \phi_{\text{flat}}^{-}(t')\phi_{\text{flat}}^{+}(t)-\phi_{\text{flat}}^{+}(t)\phi_{\text{flat}}^{+}(t') \frac{\phi_{\text{flat}}^{-}(t_{0})}{\phi_{\text{flat}}^{+}(t_{0})}\right]\,,\\
&= \frac{i}{2k}\left[e^{-ik(t-t')}\theta(t-t')+e^{-ik(t'-t)}\theta(t'-t)-e^{ik(t+t')} \right]	\,.
\end{align}
Since $  K $ takes such a simple form, the integral representation of the wavefunction can also be written as 
\begin{equation}\label{seed}
	 \psiflat(\{k_n\}, \{ p_m\}) = i\int d^V t e^{i q\cdot  t} \prod^{I}_m  G_{\text{flat}}( p_m)\,,
\end{equation}
where dot products of momenta and conformal time are taken to mean 
\begin{align}
q\cdot t = \sum_{A=1}^{V} q_A t_A\,,
\end{align} 
and $  q_{A} $ is the sum of all external energies ending on the $  A $ vertex, $q_A = \sum k_a$ for $  a \in A $.
Note that the dependence of $  \psiflat $ on the boundary energies $  k_{n} $ is only through the vertex energies $q_A$ at each vertex. Because of this, in Steps 1 and Step 2 we do not need to know how many external lines are connected to a vertex, only the sum of those external energies appears. 

The exchange energies $p_m$ are uniquely fixed by imposing \textit{momentum} conservation at every vertex. Notice however that ``energy'', i.e. the norm $  k=|\v{k}| $ of the three-vectors $  \v{k} $ is not conserved. 

The seed functions $  \psiflat_{n} $ can of course be computed explicitly expanding the products of bulk-to-bulk propagators and computing the time integrals in \eqref{seed}, but there are more efficient methods as described in \cite{cosmopoly}.  Here we review the presentation in \cite{cosmopoly} of the Old-Fashioned Perturbation Theory (OFPT) expansion.\\
\indent Consider 
\begin{equation}
	\left(\sum\limits_A  q_A \right)\psiflat = i\int d^V\eta\left(\Delta_\eta e^{i q\cdot  \eta}\right)\prod\limits_m G_{\text{flat}}( p_m)\,,
\end{equation}
where we have inserted the time translation generator
\begin{align*}
	\Delta_\eta = -i\sum\limits_a \partial_{\eta_a}
\end{align*}
We can integrate by parts to obtain 
\begin{equation}
	\left(\sum\limits_A  q_A \right)\psiflat = -i\int d^V\eta e^{iq\cdot \eta}\sum\limits_m \Delta_\eta G_{\text{flat}}(\eta_{m}, \eta_{m}', p_m)\prod\limits_{m' \neq m}G_{\text{flat}}(\eta_{m'}, \eta_{m'}', p_{m'})
\end{equation}
We note that the time-ordered part of the propagator $G_{\text{flat}}$ is time-translation invariant and therefore annihilated by $\Delta_\eta$ so that 
\begin{equation}
\label{edgedel}
	\Delta_\eta G_{\text{flat}}(\eta_{m}, \eta_{m}', p_m) = -ie^{ip_m(\eta_{m}+ \eta_{m}')}\,.
\end{equation}
This has the effect of deleting the propagator and shifting the energies at the two relevant vertices.  From this we derive the expression 
\begin{equation}\label{graphicalrep}
	\begin{tikzpicture}[baseline=(current  bounding  box.center)]
		\coordinate (psi) at (-4, 0);
		\coordinate (psiL) at (-0.5, 0);
		\coordinate (psiR) at (2, 0);
		\coordinate (loop) at (5, 0);
		\draw[thick] (psi) circle (7mm);
		\draw[thick] (psiL) circle (7mm);
		\draw[thick] (psiR) circle (7mm);
		\node at (-2.65, 0) {$=$};
		\node at (-5.5, 0) {$\left(\sum\limits_A q_A \right)$};
		\node at (-4, 0) {$\large\psi$};
		\node at (psiL) {$\psi_L^{+p_m}$};
		\node at (2, 0) {$\psi_R^{+p_m}$};
		\draw[dashed] (.2, 0) -- (1.3, 0);
		\node at (.75, -.2) {$m$};
		\node at (-1.75, -.2) {\LARGE $\sum\limits_m $};
		\node at (3.5, 0) {$+$};
		\draw[color=red] (loop) circle (7mm);
		\node at (5, 0) {$\color{red}\psi_{l-1}$};
		\draw[dashed, color=red] (4.5, -.5) to [out=240, in=180] (5, -1.5) to [out=0, in=310] (5.5, -.5);
		\node at (5, -1.75) {$\color{red}m$};
		\node at (4.2, -.5) {\tiny $\color{red}+p_m$};
		\node at (5.85, -.5) {\tiny $\color{red} +p_m$};
	\end{tikzpicture}	\,,
\end{equation}
\text{}\\
where the sum runs over all deletions of a single edge $m$.  We can ignore the term colored in red since we have restricted our analysis to tree level.  The relevant term is therefore the first graphic corresponding to edges that disconnect the diagram upon deletion.  The vertices connected by the deleted edge now absorb positive edge energy.  This is made clear in \eqref{edgedel}.  From the contact seed which is simply $1/p$ we can apply this recursive expression to efficiently build any seed function $\psiflat$ defined by the the truncated diagram with associated edge and vertex energies. As discussed in \cite{Cespedes:2020xqq}, the algebraic nature of the above recursion relations can be traced back to the fact that in time-translation invariant theories the Schr\"odinger equation becomes algebraic. 


\subsection{Examples}

Here we illustrate the recursive calculation of $  \psiflat $ in Step 1 with some examples.


\paragraph{Two-site chain} First we note that the base case is 
\begin{equation}
\psiflat_{1}=\bullet_{q}\hspace{3mm} = \hspace{3mm} \frac{1}{q},
\end{equation}
where $ q $ represents the sum of all the energies of the external legs attached to this vertex. Since it is only this sum that appears in all expression, when representing a diagram we don't show the external legs. For this contact interaction the diagram reduces to a single dot. 
Now we can consider the single factorization channel determining the two-site chain with intermediate energy $  p $
\begin{equation}
\begin{tikzpicture}[baseline=(current  bounding  box.center)]
\node at (-4, 0) {$\left(q_1+q_2 \right)$};
			\node at (-3, 0) {$\bullet $};
			\node at (-1, 0) {$\bullet$};
			\draw (-3, 0) -- (-1, 0);
			\node at (0, 0) {$=$};
			\node at (1, 0) {$\bullet$}; 
			\draw (1, 0) -- (1.5, 0);
			\draw (2.5, 0) -- (3, 0);
			\node at (3, 0) {$\bullet$};
			\draw[red] (2, .35) -- (2, -.25);
			\node at (1, -.35) {$q_1$};
			\node at (2.75, -.25) {$q_2$};
			\node at (1.5, .25) {$+p$};
			\node at (2.5, .25) {$+p$};
			\node at (-2, -.25) {$p$};
\end{tikzpicture}\,.
\end{equation}
This tells us that the wavefunction coefficient is 
\begin{equation} \label{twosite}
\psiflat_{2} = \frac{\psiflat_{1}(q_{1}+p)\psiflat_{1}(q_{2}+p)}{q_{T}}=\frac{1}{(q_1+q_2)(q_1+p)(q_2+p)}\,.
\end{equation}


\paragraph{Three-site chain}
Now we move on to the three-site chain and comment on the general structure of the recursion graphically. Every term in the recursion is a sequence of cut edges and every cut corresponds to a product of contributions from the diagrams on either side of the cut.  Every sequence terminates with all edges cut and therefore we get the inverse product of vertex energy $  q $ plus the energy of all edges landing on it, summed over all vertices. This motivates the circling depiction of contributions to the recursive sum.\\
  At each step we circle the two disconnected subgraphs produced by the deletion of a given edge.  We then proceed within each subgraph until we reach individual vertices.  Each circling furnishes a factor of the inverse of the sum of encircled vertex energies plus cut edge energies.  Below we depict the two-site chain computed in the previous example:
\begin{equation}
  		\begin{tikzpicture}[baseline=(current  bounding  box.center)]
  	        \coordinate (x1) at (-4, 0);
  	        \coordinate (x2) at (-2, 0);
  	        \coordinate (v1) at (-4.2, 0);
  	        \coordinate (v2) at (-1.8, 0);
  			\node at (x1) {$\bullet$};
  			\node at (x2) {$\bullet$};
  			\draw (x1) -- (x2);
  			\draw[color = blue] (v1) to [out=90, in = 180] (-3, .475) to [out = 0, in = 90] (v2) to [out= 270, in = 0] (-3, -.455) to [out = 180, in=270] (v1);
  			\draw[color = red] (x1) circle (1.5mm);
  			\draw[color = red] (x2) circle (1.5mm);
  			\node at (-3, -.22) {$p$};
  		\end{tikzpicture}	\quad 
\psiflat_{2}= \frac{{\color{red}\psiflat_{1}(q_{1}+p) \psiflat_{1}(q_{2}+p)}}{\color{blue} \sum_{A}^{2} q_{A}} =  \frac{1}{q_1+q_2}  \frac{1}{q_1+p}\frac{1}{q_2+p}\,.
\end{equation}
 Again, noting that each such contribution will contain the same total energy factor and individual vertex factors, which can be pulled out in front for any computation.  Then what is left to be considered is all sequences of proper subgraphs with at least two vertices.  We see this with the three-site chain.
 \begin{align}\label{psiflat3}
 		\begin{tikzpicture}[baseline=(current  bounding  box.center)]
 			\coordinate (x1) at (-6, 0);
 			\coordinate (x2) at (-4, 0);
 			\coordinate (x3) at (-2, 0);
 			\coordinate (v1) at (-6.2, 0);
 			\coordinate (v2) at (-3.8, 0);
 			\coordinate (v3) at (-4.2, 0);
 			\coordinate (v4) at (-1.8, 0);
 			\node at (x1) {$\bullet$};
 			\node at (x2) {$\bullet$};
 			\node at (x3) {$\bullet$};
 			\draw (x1) -- (x2) -- (x3);
 			\draw[color = violet] (v1) to [out=90, in = 180] (-5, .475) to [out = 0, in = 90] (v2) to [out= 270, in = 0] (-5, -.455) to [out = 180, in=270] (v1);
 			\draw[color = orange] (v3) to [out=90, in = 180] (-3, .475) to [out = 0, in = 90] (v4) to [out= 270, in = 0] (-3, -.455) to [out = 180, in=270] (v3);
 			\draw[color = orange] (x1) circle (1.5mm);
  			\draw[color = violet] (x3) circle (1.5mm);
 			\node at (-5, -.22) {$p_1$};
 			\node at (-3, -.22) {$p_2$};
 		\end{tikzpicture}
\end{align}
\begin{align}
\psiflat_{3}&=\frac{{\color{violet} \psiflat_{2}(q_{1},q_{2}+p_{2})\psiflat_{1}(q_{3}+p_{2})}+{\color{orange}\psiflat_{1}(q_{1}+p_{1})\psiflat_{2}(q_{2}+p_{1},q_{3})}}{ \sum_{A}^{3} q_{A}} \nonumber \\
&=\frac{1}{(q_1+q_2+q_3)(q_1+p_1)(q_2+p_1+p_2)(q_3+p_2)}\left( \frac{1}{q_1+q_2+p_2}+ \frac{1}{q_2+q_3+p_1} \right)\,,
\end{align}
 Where we depicted a sum over two choices of proper subgraph (equivalently, single edge deletions).  The initial circling as well as the subsequent and complementary vertex circlings are omitted as their factors have been accounted for out front. 
 
 
 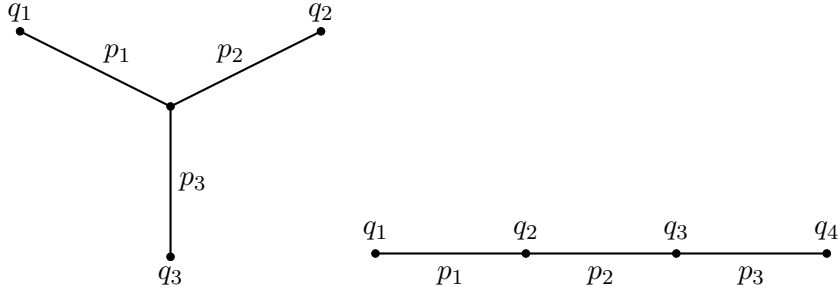
\begin{figure}
 \centering
 \begin{tikzpicture}
			\coordinate (v1) at (-2, 1);
			\coordinate (v2) at (2,1);
			\coordinate (v3) at (0,-2);
			\coordinate (v4) at (0,0);
			\draw[thick] (v1) -- (v4);
			\draw[thick] (v2) -- (v4);
			\draw[thick] (v3) -- (v4);			
			\draw[fill] (v1) circle (.5mm) node[anchor=south]{$  q_{1} $};
			\draw[fill] (v2) circle (.5mm) node[anchor=south]{$  q_{2}$};
			\draw[fill] (v3) circle (.5mm) node[anchor=north]{$  q_{3} $};
			\draw[fill] (v4) circle (.5mm) ;
			\node at (-.7, .7) {$p_{1}$};
			\node at (.8, .7) {$p_{2}$};
			\node at (0.3, -1) {$p_{3}$};
\end{tikzpicture} 
\begin{tikzpicture}
			\coordinate (v1) at (-2, 0);
			\coordinate (v2) at (0, 0);
			\coordinate (v3) at (2, 0);
			\coordinate (v4) at (4, 0);
			\draw[thick] (v1) -- (v2) --(v3) -- (v4);
			\draw[fill] (v1) circle (.5mm);
			\draw[fill] (v2) circle (.5mm);
			\draw[fill] (v3) circle (.5mm);
			\draw[fill] (v4) circle (.5mm);
			\node at (-2,0.3) {$  q_{1} $};
			\node at (-1,-0.3) {$  p_{1} $};
			\node at (0,0.3) {$  q_{2} $};
			\node at (1,-0.3) {$  p_{2} $};
			\node at (2,0.3) {$  q_{3} $};
			\node at (3,-0.3) {$  p_{3} $};
			\node at (4,0.3) {$  q_{4} $};
		\end{tikzpicture}
 \caption{The two topologies for the four-site skeleton diagrams, namely the flux capacitor (left) and the four-site chain (right). The diagram also indicates the total external energies $  q_{A} $ associated with each vertex $  A=1,\dots,4 $ and the energy $  p_{m} $ of each internal line (edge) $  m=1,2,3 $.\label{4}}
 \end{figure}
 
 \paragraph{Four-site diagrams: the flux capacitor} As a last example we discuss the two possible topologies for the four-site diagrams, namely the flux capacitor and the four-site chain. For the flux capacitor, with the kinematical assignments given in Figure \ref{4}, we find:
 \begin{align}
\psiflat_{4}= \frac{  \psiflat_{3}(q_{1},q_{4}+p_{3},q_{2}) \psiflat_{1}(p_{3}+q_{3})+ \text{2 perm's}
}{q_{1}+q_{2}+q_{3}+q_{4}} \quad \quad \text{(flux capacitor)}\,.
 \end{align}
For the four-site chain, with kinematics given in Figure \ref{4}, we find the seed function
\begin{align}
\psiflat_{4}&= \frac{1}{q_{1}+q_{2}+q_{3}+q_{4}} \left[ \psiflat_{3}(q_{1},q_{2},q_{3}+p_{3}) \psiflat_{1}(p_{3}+q_{4}) \right. \\ 
& \quad \left. + \psiflat_{1}(q_{1}+p_{1}) \psiflat_{3}(p_{1}+q_{2},q_{3},q_{4})+ \psiflat_{2}(q_{1},q_{2}+p_{2})\psiflat_{2}(p_{2}+q_{3},q_{4})\right] \quad  \text{(four-site chain)}\,. \nonumber
 \end{align}
The general procedure is straightforward, efficient, and readily furnishes the necessary seed functions with only algebraic manipulations.

 
 \section{Step 2: The intermediate wavefunction $\psint$}\label{sec:4}
 
 In this section, we prescribe differential operators that transform the flat spacetime wavefunction $  \psiflat $ of Step 1 into an intermediate wavefunction, $  \psint $, with the correct exchange propagators for a massless scalar field in dS up to overall powers of conformal time in the integrand. All the additional time dependence brought about from the vertices and the associate powers of spatial derivatives will be included in Step 3, which is discussed in the next section.
 
Before beginning the derivation let's briefly introduce our notation. For scalars, the bulk-to-boundary propagator $  K $ is given in terms of the mode functions $  \phi^{+} $ by 
\begin{align}\label{masslessmodefct}
K(k,\eta)&=\frac{\phi^{+}(k,\eta)}{\phi^{+}(k,0)}= \frac{\frac{H}{\sqrt{2k^{3}}}\left(  1-ik\eta\right) e^{ik\eta} }{\frac{H}{\sqrt{2k^{3}}}\left(  1-ik0\right) e^{ik0} }=\left(  1-ik\eta\right) e^{ik\eta}\,,
\end{align} 
where we took $  \eta_{0}\to 0 $. The bulk-to-boundary propagator is 
\begin{align}\label{G}
G(\eta, \eta', k) &=iP(k)\left[  \theta(\eta-\eta') K^{\ast} \nonumber(\eta)K(\eta')+\theta(\eta'-\eta) K^{\ast}(\eta')K(\eta)-K(\eta)K(\eta')\right] \\
&=i\left[  \theta(t-t') \phi^{-}(t)\phi^{+}(t')+\theta(t'-t) \phi^{-}(t')\phi^{+}(t)-\phi^{+}(t)\phi^{+}(t') \frac{\phi^{-}(t_{0})}{\phi^{+}(t_{0})}\right]\,,
\end{align}
where now the power spectrum for a canonically normalized massless scalar in de Sitter is $  P(k)=H^{2}/(2k^{3}) $ and we set $  H=1 $ throughout. Notice that $\partial_{t} K$ carries two powers of conformal time 
\begin{align}
\partial_{t}K(k,\eta)=\frac{1}{a}\partial_{\eta}K(k,\eta)=- H k^{2}\eta^{2} e^{ik\eta} \,.
\end{align}
We denote derivatives with respect to $ t  $ with a dot and  those with respect to $\eta$ with a prime. The above propagators and mode functions are invariant under the full dS isometries, including dS boosts. Conversely, we will allow for interactions that break dS boosts. This is important to capture many of the leading phenomenological models of inflation, where the breaking of dS boost is not slow-roll suppressed but can be large. In fact, for scalar perturbations in single-field inflation the breaking of boosts is a necessary condition to have non-vanishing connected correlators \cite{Green:2020ebl}.

For simplicity, we will only consider parity even interactions, but we expect the generalization to the parity odd case to be straightforward. Assuming parity, every pair of spatial derivatives $  \partial_{i} $ needs to be contracted with the inverse metric $  g^{ij}=\delta_{ij} \eta^{2}H^{2} $,  and so every spatial derivative carries one power of conformal time. The only inverse powers of conformal time come from $  \sqrt{-g}=(H\eta)^{-4} $ in the measure of the covariant spacetime integral (in conformal time). For example, a tree-level contribution to a de Sitter wavefunction coefficient $ \psi_{n} $ represented by a tree diagram with $ I$ internal lines and $ V$ vertices without any time derivatives takes the form
\begin{equation}\label{seed}
	 \psi(\{k_n\}; \{ p_m\};\{\v{k}\}) = i\int \left[  \prod_{A=1}^{V} d\eta_{A} F_{A} \prod_{a\in A}K(k_{a},\eta_{A})\right] \left[ \prod\limits_{m=1}^{I}  G( p_m) \right]\,,
\end{equation}
where each conformal time $\eta_A$ is integrated from $  -\infty(1-i\e) $ to $  0 $ with $  \e>0 $ being taken to zero at the end of the calculation.  This prescription projects out the Bunch-Davies vacuum in the past. Here, $ F_{A}$ represents the vertex factor associated to each vertex $ A$ and collects all contractions of spatial momenta and polarization tensors, as well as all the associated powers of $ \eta$. Where no confusion arises, we will omit specifying the dependence of $ \psi_{n}$ on the internal energies $ p_{m}$ and the scalar products of momenta $ \{ \v{k}\}$. In the presence of time derivative interactions some of the bulk-to-boundary propagators $ K$ may become $ K'$ and some of the bulk-to-bulk propagators $ G=G_{\phi\phi}$ may become $ G_{\phi'\phi}$, $ G_{\phi\phi'}$ or $ G_{\phi' \phi'}$.
 
 
 \subsection{Motivation}
 Before stating the general prescription for the intermediate wavefunction, we motivate it with an explicit example.  Consider the time integral
  \begin{align*}
  	\psi= i\int d\eta_1 d\eta_2 d\eta_3 \,e^{i \sum_{A}^{3}q_{A} \eta_{A}}\,G_{\phi' \phi'}(\eta_1, \eta_2, p_1)G_{\phi' \phi}(\eta_2, \eta_3, p_2)\,.
  \end{align*}   
  Using \eqref{propspp} and \eqref{propsp} and ignoring overall powers of conformal time we are motivated to consider the function
 \begin{align*}
 	\psint = i\int d\eta_1 d\eta_2 d\eta_3 \, e^{i \sum_{A}^{3}q_{A} \eta_{A}}\left[p_1^2 G_{\text{flat}}(\eta_1, \eta_2, p_1)-\delta(\eta_1-\eta_2) \right](1-\eta_3\partial_{\eta_3})G_{\text{flat}}(\eta_2, \eta_3, p_2)\,.
 \end{align*}
The quantities $\psi_{\text{int}}$ and $\psi$ differ by overall powers of the conformal times $ \eta_{1,2,3}$. However, this difference can be fixed by taking an appropriate number of derivatives with respect to $q_A$, which act exclusively on the exponential bringing down the desired powers of $ \eta_{1,2,3}$. This is something we will do in Step 3 and so, for the moment, we can neglect this difference. Our objective here is instead to express $ \psint$ as an operator acting on the three-site chain seed $  \psiflat_{3} $ and the two-site chain seed $  \psiflat_{2} $ (because of the $\delta$-function term). Integrating by parts in $\eta_3$, it is easy to see that the answer is 
 \begin{align*}
 	\psint = 	\left( 
 	2+q_3\partial_{q_3}\right)\left[p_1^2\psiflat_3-\psiflat_2\right]\,.
 \end{align*}
 Via the application of external derivatives and contractions of polarization tensors and momenta, arbitrarily many external particles with arbitrary interactions can be attached to the three vertices.  Therefore $\psint$  corresponds to intermediate building block for infinite classes of wave-function coefficients for massless scalars and gravitons. Let's now move on to consider arbitrary trees.


 \subsection{The general case}
 
Recall the relations we found between derivatives of the de Sitter bulk-to-bulk propagator $ G$ and the flat space propagator $ G_{\text{flat}}$, \eqref{props}-\eqref{propsp}, which we report here for convenience
\begin{align}\label{props2}
G_{\phi\phi}(\eta, \eta', p) &= \frac{1}{p^2}\left[ (1-\eta\partial_\eta)(1-\eta'\partial_{\eta'})G_{\text{flat}}(\eta, \eta', p)+\eta\eta'\delta(\eta-\eta')\right]\,,\\
G_{\phi'\phi'}(\eta, \eta', p) &= \eta\eta'\left[p^2G_{\text{flat}}(\eta, \eta', p)-\delta(\eta-\eta') \right]\,,\label{propspp2} \\
G_{\phi \phi'}(\eta, \eta', p) &= (1-\eta\partial_\eta)\eta' G_{\text{flat}}(\eta, \eta', p) =G_{\phi' \phi}(\eta', \eta, p) \,.\label{propsp2}
\end{align}
When all bulk-to-bulk propagators $ G$ for a given diagram have been re-written as above, we would like to interpret the resulting expression as a differential operator acting on a flatspace seed wavefunction $ \psiflat$. To do this, we have to overcome two obstacles. First we have to account for the factors $(1-\eta \partial_\eta)$. We will now show that, by repeated use of integration by parts in time, we can trade these time derivatives for differential operators that act exclusively on external kinematics. We refer to this procedure as \textit{re-routing}. Second, because of the delta functions appearing above, the corresponding term can be obtained by acting on a flatspace wavefunction $ \psiflat$ with the relevant edge collapsed. In particular, for a generic diagram with a number $E_{\phi\phi}$ of $ G_{\phi\phi}$ propagators and a number $E_{ \phi'  \phi'}$ of $ G_{\phi'\phi'}$ propagators, we must sum over all combinations of collapsing $\braket{ \phi'  \phi'}$ and $\braket{\phi\phi}$ edges. This produces a sum over $2^{E_{ \phi'  \phi'}+E_{\phi\phi}}$ diagrams with only propagators and no $\delta$-functions. We refer to this procedure as \textit{collapsing}.

 
\paragraph{On re-routing} Given a time integral with a factor of $(1-\eta \partial_\eta)$ for every half-edge that is labeled by a $\phi$, the objective is to re-route the time derivatives to a vertex that is connected only to a single exchange edge.  Then a final integration by parts acts purely on external mode functions and can therefore be traded for a derivative with respect to some external energies $ q$.
 \begin{figure}
 	\begin{center}
 		\begin{tikzpicture}[scale=1.1]
 			\coordinate (v1) at (-2, 1.7);
 			\coordinate (v2) at (2, 1.7);
 			\coordinate (v3) at (0, -2.1);
 			\coordinate (v4) at (0, 0);
 			\draw[thick] (v1) -- (v4) -- (v3);
 			\draw[thick] (v4) -- (v2);
 			\draw[fill] (v1) circle (.5mm);
 			\draw[fill] (v2) circle (.5mm);
 			\draw[fill] (v3) circle (.5mm);
 			\draw[fill] (v4) circle (.5mm);
 			\node at (1.6, 1.7) {$\phi$};
 			\node at (-1.6, 1.7) {$\dot \phi$};
 			\node at (.2, .45) {$\phi$};
 			\node at (.2, -.3) {$\dot \phi$};
 			\node at (-.4, 0) {$\phi$};
 			\node at (.2, -1.8) {$\dot \phi$};
 			\draw[->] (-.6, 0.1) -- (-1.2, .6);
 			\draw[->] (1.2, 1.8) -- (1.8, 2.3);
 			\draw[->] (0, .7) -- (.6, 1.2);
 		\end{tikzpicture}
 	\end{center}
 	\caption{Edge-labeled diagram with a choice of routing for the two $\phi$-type half-edges attached to the central vertex.}
 \end{figure}
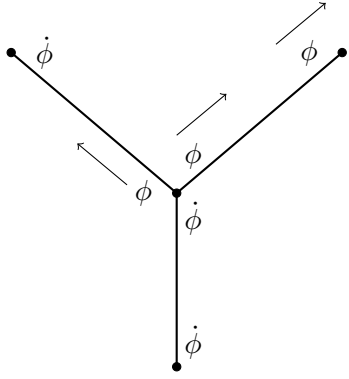
 
 \noindent To begin, we note the relation 
 \begin{equation}\label{vici}
  (1+p\partial_p -\eta_{B} \partial_{\eta_{B}}-	\eta_{A} \partial_{\eta_{A}})G_{\text{flat}}(\eta_{A}, \eta_{B}, p)=0\,.
 \end{equation} 
Given an edge that connects vertices at time $  \eta_{A}$ and $  \eta_{B} $, consider the half-edge associated with time $\eta_{A}$.  Using \eqref{vici}, we can choose to flow in one of two directions: either through the vertex at time $\eta_{A}$ or through the vertex at time $\eta_{B}$. This choice is depicted in Figure \ref{routing}.  In each case, integration by parts forces us to accumulate $(1+p_{m}\partial_{p_{m}})$ for every edge $m$ we flow through and $(1+q_{C}\partial_{q_{C}})$ for every vertex $C$ we flow through, with a relative sign between the two choices of direction.  In this way we arrive at the following differential operators associated to each $ \phi$-type half-edge 
\begin{align}\label{halfedgediff}
&\text{Re-routing vertex side:}&\Drr& = 1+\left[\sum\limits_C \left(1+q_C\partial_{q_C}\right)+\sum\limits_m \left( 1+p_m\partial_{p_m} \right) \right] \,,\\ 
&\text{Re-routing opposite-vertex side:}&\Drr &= 1-\left[\sum\limits_C \left(1+q_C\partial_{q_C}\right)+\sum\limits_m \left( 1+p_m\partial_{p_m} \right)\right] \,,\hspace{3mm}
\end{align}
where the sums run over all vertices\footnote{In our procedure \textit{every} vertex $ C$ has some corresponding $ q_{C}$, even if in the final graph there are no external legs attached to that vertex. This vertex energy $ q_{C}$ is hit by various $ \partial_{q_{C}}$ derivatives, as for example here in the re-routing operator or in \eqref{etapower} in Step 3. It is only \textit{after} all derivatives have been taken that the  $ q_{C}$ corresponding to a given vertex should be set to zero if there are no external lines attached to it.} $ C$ and edges $ m$ encountered as one flows from the given half-edge out of diagram. For each $ \phi$-type half edge, there are two choices for $ \Drr$ corresponding to flowing in the direction of the vertex to which the half-edge is attached, namely ``vertex side'', or in the opposite direction, namely ``opposite vertex side''. The rerouting procedure prescribed above relies upon the tree approximation.

\paragraph{On collapsing edges:} To account for the delta functions in \eqref{props2} and \eqref{propspp2} we have to consider diagrams where one or more edges have been collapsed and the associated vertices have been merged. The re-routing operator in \eqref{halfedgediff} is to be applied for each $\phi$ half-edge for each of the diagrams in the set of $2^{E_{ \phi' \phi'}+E_{\phi\phi}}$ ways of collapsing edges.  We now provide some notation for organizing the collapsing of edges. It is important to distinguish between the the collapsing of $ \phi\phi$ or $ \phi'\phi'$ edges as they have a different factors. The differential operators we will discuss in Step 3 will act the sum of all collapsed contributions. 
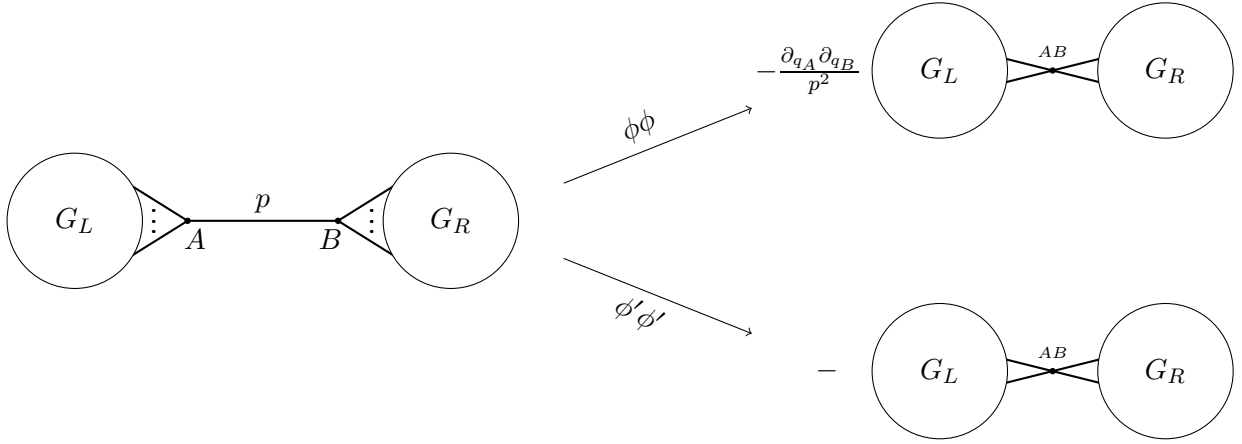
\begin{figure}
	\centering
	\begin{tikzpicture}[scale=.5]
		\coordinate (v1) at (-2, 0);
		\coordinate (v2) at (2, 0);
		\draw[thick] (v1) -- (v2);
		\draw[thick] (v1) -- (-4, 1.25);
		\draw[thick] (v1) -- (-4, -1.25);
		\draw[thick] (v2) -- (4, 1.25);
		\draw[thick] (v2) -- (4, -1.25);
		\draw[fill=white] (-5, 0) circle(18mm);
		\draw[fill=white] (5, 0) circle(18mm);
		\draw[fill] (v1) circle (.7mm);
		\draw[fill] (v2) circle (.7mm);
		\node at (-2.9, .2) {$\vdots$};
		\node at (2.9, .2) {$\vdots$};
		\node at (-1.8, -.5) {$A$};
		\node at (1.8, -.5) {$B$};
		\node at (-5, 0) {$G_L$};
		\node at (5, 0) {$G_R$};
		\draw[->] (8, 1) -- (13, 3);
		\draw[->] (8, -1) -- (13, -3);
		\node[rotate=27] at (10, 2.5) {$\phi\phi$}; 
		\node[rotate=-27] at (10, -2.5) {$ \phi' \phi'$}; 
		\node at (0, .45) {$p$};
		\coordinate (w1) at (21, 4);
		\draw[thick] (w1) -- (17, 5);
		\draw[thick] (w1) -- (17, 3);
		\draw[thick] (w1) -- (25, 5);
		\draw[thick] (w1) -- (25, 3);
		\draw[fill=white] (18, 4) circle(18mm);
		\draw[fill=white] (24, 4) circle(18mm);
		\draw[fill] (w1) circle (.7mm);
		\node at (21, 4.5) {\tiny $AB$ };
		\node at (-2.9, .2) {$\vdots$};
		\node at (2.9, .2) {$\vdots$};
		\node at (18, 4) {$G_L$};
		\node at (24, 4) {$G_R$};
		\coordinate (z1) at (21, -4);
		\draw[thick] (z1) -- (17, -5);
		\draw[thick] (z1) -- (17, -3);
		\draw[thick] (z1) -- (25, -5);
		\draw[thick] (z1) -- (25, -3);
		\draw[fill=white] (18, -4) circle(18mm);
		\draw[fill=white] (24, -4) circle(18mm);
		\node at (18, -4) {$G_L$};
		\node at (24, -4) {$G_R$};
		\draw[fill] (21, -4) circle (.7mm);
		\node at (21, -3.5) {\tiny $AB$ };
		\node at (21, -3.5) {\tiny };
		\node at (21, -5) {\tiny };
		\node at (15, -4) {$-$};
		\node at (14.5, 4) {$-\frac{\partial_{q_{A}}\partial_{q_{B}}}{p^{2}}$};
	\end{tikzpicture}
	\caption{Here we depict edge collapsing in the case of $\phi\phi$ and $\phi' \phi'$ propagators. For $\phi\phi$ we need a factor of $p^{-2}\partial_{q_{A}}\partial_{q_{B}} $ whereas for $\phi'\phi'$ just a factor of $-1$. }
	\label{collapsing}
\end{figure}
Moreover, when both collapsing and re-routing are needed, the necessary re-routing operators $\Drr$ need to be applied to each collapsed contribution. Here we specify what data is necessary to associate with the merged vertex produced upon collapsing an edge.  Figure \ref{collapsing} depicts the two possibilities for collapsing.  We consider an edge connecting vertices $ A$ and $B$ collapsing down to an effective vertex $AB$.  Denote as $ d_A$ and $ d_B$ the powers of conformal time at the respective vertices, which are given by \eqref{bound}. We find in each case
\begin{align}\label{collapsedeq}
	\text{collapsing } \phi\phi &\to \hspace{3mm} \eta_{A}^{d_A-4 }\eta_{B}^{d_{B
	}-4}\left[\frac{1}{p^2} \eta_A\eta_{B}\delta(\eta_A-\eta_{B})\right]\,,\\ 
	\text{collapsing } \phi'\phi' &\to  \hspace{3mm} \eta_A^{d_{A}-4 }\eta_{B}^{d_{B}-4}\left[-\delta(\eta_A-\eta_{B})\right]\,. \label{collapsedeq2}
\end{align}
Since the overall powers of $ \eta$ will be accounted for in Step 3, the factors that need to be included for each collapsed edge in Step 2 are simply
\begin{align}\label{collapseop}
\text{collapsing } \phi\phi &\to-	\frac{1}{p^2}\partial_{q_{A}}\partial_{q_{B}}  & \text{collapsing } \phi'\phi' &\to	-1 \,.
\end{align}
The above operators should be applied after the application of re-routing operators on the collapsed diagram. Finally, the overall powers of $ p_{m}$ in \eqref{props2}-\eqref{propsp2} require that we multiply all terms contributing in Step 2 by the overall factor
\begin{equation}
\label{Pfact}
\text{P} = 	\prod\limits_{\phi'\phi'}p_m^2 \prod\limits_{\phi\phi}\frac{1}{p_m^2}\,,
\end{equation}
where the products are over propagators of each type.

In summary, for each internal propagator in Step 2 we have to perform the following operations
\begin{align}\label{summar}
\phi' \phi' &\to \text{collapsing}\,,&
\phi \phi' &\to \text{re-routing}\,,&
\phi \phi &\to \text{collapsing \& re-routing} \,.
\end{align}


\subsection{Examples}\label{step2ex}

In the following we present several examples with two, three and four vertices (``sites''), respectively. 


\paragraph{Two-site classification:} Now we classify the different $\psint$ with a two-site skeleton.  There are three possibilities, corresponding to the three choices for the bulk-to-bulk propagator (up to permutations). The $  \phi'\phi' $ case requires only collapsing an edge according to \eqref{collapsedeq2}, but no re-routing. Since all overall factors of $  \eta $ will be included in Step 3, we can neglected them at this stage, and the result is simply
\begin{equation}\label{phipphip}
	\begin{tikzpicture}[baseline=(current  bounding  box.center)]
		\coordinate (v1) at (-6, 0);
		\coordinate (v2) at (-2, 0);
		\draw[thick] (v1) -- (v2);
		\draw[fill] (v1) circle (.5mm);
		\draw[fill] (v2) circle (.5mm);
		\node at (-5.75, .3) {$ \phi'$};
		\node at (-2.25, .3) {$ \phi'$};
		\node at (3, 0) {$\psint^{\phi'\phi'}= p^{2} \psiflat_{2}(q_{1},q_{2})-\psiflat_{1}(q_{1}+q_{2}) $,};
		\node at (-6, -.3) {$q_1$};
		\node at (-2, -.3) {$q_2$};
		\node at (-4, -.3) {$p$};
	\end{tikzpicture}
\end{equation}
where the arguments of $  \psiflat $ refer to the sums $  q_{1} $ and $  q_{2} $ of external energies connected to each of the two vertices. For example, $  \psi_{\text{flat},1}  $ has a single external energy which should be taken to be the sum of the external energies at the two vertices that have been collapsed to one by the $  \delta $ function in \eqref{propspp}. So it's argument is $  q_{1}+q_{2} $. Notice the factor of $ -1$ in front of the contribution from collapsing the $ \phi'\phi'$ edge, as dictated by \eqref{collapseop}.

The $  \phi'\phi $ case requires re-routing with the operators in \eqref{halfedgediff}, but no collapsing is needed since there are no delta functions in \eqref{propspp}. Re-routing the factor of $  (1-\eta' \partial_{\eta'}) $ on the $  \phi $ vertex side using the vertex-side operator in \eqref{halfedgediff} we find
\begin{equation}\label{phipphidraw}
	\begin{tikzpicture}[baseline=(current  bounding  box.center)]
		\coordinate (v1) at (-6, 0);
		\coordinate (v2) at (-2, 0);
		\draw[thick] (v1) -- (v2);
		\draw[fill] (v1) circle (.5mm);
		\draw[fill] (v2) circle (.5mm);
		\node at (-5.75, .3) {$ \phi'$};
		\node at (-2.25, .3) {$ \phi$};
		\node at (3, 0) {$\psint^{\phi'\phi}= (2+q_2\partial_{q_2}) \psiflat_{2}(q_{1},q_{2})$.};
		\node at (-6, -.3) {$q_1$};
		\node at (-2, -.3) {$q_2$};
		\node at (-4, -.3) {$p$};
	\end{tikzpicture}
\end{equation}
The propagator $  G_{\phi\phi} $ was given in \eqref{props}. When re-written in terms of $ G_{\text{flat}}$ it has both a delta function and time derivatives so it needs both collapsing and re-routing (see \eqref{summar}). Re-routing vertex-side each of the time derivatives with \eqref{halfedgediff} we find the final result
\begin{equation}\label{phiphidraw}
	\begin{tikzpicture}[baseline=(current  bounding  box.center)]
		\coordinate (v1) at (-6, 0);
		\coordinate (v2) at (-2, 0);
		\draw[thick] (v1) -- (v2);
		\draw[fill] (v1) circle (.5mm);
		\draw[fill] (v2) circle (.5mm);
		\node at (-5.75, .3) {$\phi$};
		\node at (-2.25, .3) {$\phi$};
		\node at (4, 0) {$\psint^{\phi\phi}=\frac{1}{p^2} (2+q_1\partial_{q_1})(2+q_2\partial_{q_2}) \psiflat_2(q_{1},q_{2})-\frac{1}{p^2}\partial_{q_{1}}\partial_{q_{2}}\psiflat_1(q_{1}+q_{2}).$};
		\node at (-6, -.3) {$q_1$};
		\node at (-2, -.3) {$q_2$};
		\node at (-4, -.3) {$p$};
	\end{tikzpicture}
\end{equation}
These are the three $\psint$ needed to compute the contribution to the exchange contributions to the trispectrum from $\dot \phi^3$ and $\dot\phi (\partial\phi)^2$ interactions.  We begin to appreciate the advantages offered by our approach as we explore more complicated examples.


\paragraph{Three-site examples} The next example is the three-site chain, namely a skeleton with two bulk-to-bulk propagators and three vertices. At tree level there is a single possible topology. By appropriately collapsing and re-routing we find
\begin{align} \label{phip4}
	\begin{tikzpicture}
			\coordinate (v1) at (-2, 0);
			\coordinate (v2) at (0, 0);
			\coordinate (v3) at (2, 0);
			\draw[thick] (v1) -- (v2) --(v3);
			\draw[fill] (v1) circle (.5mm);
			\draw[fill] (v2) circle (.5mm);
			\draw[fill] (v3) circle (.5mm);
			\node at (-2,-0.3) {$  q_{1} $};
			\node at (-1,-0.3) {$  p_{1} $};
			\node at (0,-0.3) {$  q_{2} $};
			\node at (1,-0.3) {$  p_{2} $};
			\node at (2,-0.3) {$  q_{3} $};
			\node at (-1.7, .3) {$ \phi'$};
			\node at (-.4, .3) {$ \phi'$};
			\node at (.4, .3) {$ \phi'$};
			\node at (1.7, .3) {$ \phi'$};
		\end{tikzpicture} &= p_1^2 p_2^2 \psiflat_3 -p_1^2 \psiflat_{2L} -p_2^2 \psiflat_{2R}+ \psiflat_1\,,  \\ \nonumber
\begin{tikzpicture}
			\coordinate (v1) at (-2, 0);
			\coordinate (v2) at (0, 0);
			\coordinate (v3) at (2, 0);
			\draw[thick] (v1) -- (v2) --(v3);
			\draw[fill] (v1) circle (.5mm);
			\draw[fill] (v2) circle (.5mm);
			\draw[fill] (v3) circle (.5mm);
			\node at (-2,-0.3) {$  q_{1} $};
			\node at (-1,-0.3) {$  p_{1} $};
			\node at (0,-0.3) {$  q_{2} $};
			\node at (1,-0.3) {$  p_{2} $};
			\node at (2,-0.3) {$  q_{3} $};
			\node at (-1.7, .3) {$ \phi$};
			\node at (-.4, .3) {$ \phi'$};
			\node at (.4, .3) {$ \phi'$};
			\node at (1.7, .3) {$ \phi'$};
		\end{tikzpicture} &=  \left(  2+q_{1}\partial_{q_{1}}\right) \left[ p_2^2  \psiflat_3 - \psiflat_{2L} \right] , \\
\nonumber
\begin{tikzpicture}
			\coordinate (v1) at (-2, 0);
			\coordinate (v2) at (0, 0);
			\coordinate (v3) at (2, 0);
			\draw[thick] (v1) -- (v2) --(v3);
			\draw[fill] (v1) circle (.5mm);
			\draw[fill] (v2) circle (.5mm);
			\draw[fill] (v3) circle (.5mm);
			\node at (-2,-0.3) {$  q_{1} $};
			\node at (-1,-0.3) {$  p_{1} $};
			\node at (0,-0.3) {$  q_{2} $};
			\node at (1,-0.3) {$  p_{2} $};
			\node at (2,-0.3) {$  q_{3} $};
			\node at (-1.7, .3) {$ \phi$};
			\node at (-.4, .3) {$ \phi$};
			\node at (.4, .3) {$ \phi'$};
			\node at (1.7, .3) {$ \phi'$};
		\end{tikzpicture} &= \frac{1}{p_{1}^{2}} \left[\left(  2+q_{1}\partial_{q_{1}}\right) \left( -1-q_{1}\partial_{q_{1}}-p_1\partial_{p_1} \right)\psiflat_3- p_{2}^{2}\partial_{q_{1}}\partial_{q_{2}}\psiflat_{2R} \right. \\
		\nonumber & \quad \left. - \left(  2+q_{1}\partial_{q_{1}}\right) \left(  2+q_{2}\partial_{q_{2}}+q_{3}\partial_{q_{3}}\right) \psiflat_{2L} -\psiflat_{1}\right] \,,
\end{align}
as well as
\begin{align}
\nonumber
\begin{tikzpicture}
			\coordinate (v1) at (-2, 0);
			\coordinate (v2) at (0, 0);
			\coordinate (v3) at (2, 0);
			\draw[thick] (v1) -- (v2) --(v3);
			\draw[fill] (v1) circle (.5mm);
			\draw[fill] (v2) circle (.5mm);
			\draw[fill] (v3) circle (.5mm);
			\node at (-2,-0.3) {$  q_{1} $};
			\node at (-1,-0.3) {$  p_{1} $};
			\node at (0,-0.3) {$  q_{2} $};
			\node at (1,-0.3) {$  p_{2} $};
			\node at (2,-0.3) {$  q_{3} $};
			\node at (-1.7, .3) {$ \phi$};
			\node at (-.4, .3) {$ \phi'$};
			\node at (.4, .3) {$ \phi$};
			\node at (1.7, .3) {$ \phi'$};
		\end{tikzpicture} &= \left(  2+q_{1}\partial_{q_{1}}\right) \left(  -1-q_{3}\partial_{q_{3}}-p_{2}\partial_{p_{2}}\right) \psiflat_3  \,,\\
\nonumber
\begin{tikzpicture}
			\coordinate (v1) at (-2, 0);
			\coordinate (v2) at (0, 0);
			\coordinate (v3) at (2, 0);
			\draw[thick] (v1) -- (v2) --(v3);
			\draw[fill] (v1) circle (.5mm);
			\draw[fill] (v2) circle (.5mm);
			\draw[fill] (v3) circle (.5mm);
			\node at (-2,-0.3) {$  q_{1} $};
			\node at (-1,-0.3) {$  p_{1} $};
			\node at (0,-0.3) {$  q_{2} $};
			\node at (1,-0.3) {$  p_{2} $};
			\node at (2,-0.3) {$  q_{3} $};
			\node at (-1.7, .3) {$ \phi$};
			\node at (-.4, .3) {$ \phi'$};
			\node at (.4, .3) {$ \phi'$};
			\node at (1.7, .3) {$ \phi$};
		\end{tikzpicture} &= \left(  2+q_{1}\partial_{q_{1}}\right) \left(2+  q_{3}\partial_{q_{3}}\right) \psiflat_3 \,,\\
\nonumber
\begin{tikzpicture}
			\coordinate (v1) at (-2, 0);
			\coordinate (v2) at (0, 0);
			\coordinate (v3) at (2, 0);
			\draw[thick] (v1) -- (v2) --(v3);
			\draw[fill] (v1) circle (.5mm);
			\draw[fill] (v2) circle (.5mm);
			\draw[fill] (v3) circle (.5mm);
			\node at (-2,-0.3) {$  q_{1} $};
			\node at (-1,-0.3) {$  p_{1} $};
			\node at (0,-0.3) {$  q_{2} $};
			\node at (1,-0.3) {$  p_{2} $};
			\node at (2,-0.3) {$  q_{3} $};
			\node at (-1.7, .3) {$ \phi'$};
			\node at (-.4, .3) {$ \phi$};
			\node at (.4, .3) {$ \phi$};
			\node at (1.7, .3) {$ \phi'$};
		\end{tikzpicture} &= \left(  -1-p_1\partial_{p_1}-q_{1}\partial_{q_{1}}\right) \left( -1-p_2\partial_{p_2}-q_{3}\partial_{q_{3}}\right) \psiflat_3 \,,
\end{align}
and finally
\begin{align}
\begin{tikzpicture}\nonumber
			\coordinate (v1) at (-2, 0);
			\coordinate (v2) at (0, 0);
			\coordinate (v3) at (2, 0);
			\draw[thick] (v1) -- (v2) --(v3);
			\draw[fill] (v1) circle (.5mm);
			\draw[fill] (v2) circle (.5mm);
			\draw[fill] (v3) circle (.5mm);
			\node at (-2,-0.3) {$  q_{1} $};
			\node at (-1,-0.3) {$  p_{1} $};
			\node at (0,-0.3) {$  q_{2} $};
			\node at (1,-0.3) {$  p_{2} $};
			\node at (2,-0.3) {$  q_{3} $};
			\node at (-1.7, .3) {$ \phi$};
			\node at (-.4, .3) {$ \phi$};
			\node at (.4, .3) {$ \phi$};
			\node at (1.7, .3) {$ \phi'$};
		\end{tikzpicture} &=\frac{\left(  -1-q_{3}\partial_{q_{3}}-p_{2}\partial_{p_{2}}\right)}{p_{1}^{2}} \left[ \left(  2+q_{1}\partial_{q_{1}}\right) \left( -1-q_{1}\partial_{q_{1}}-p_{1}\partial_{p_{1}}\right)  \psiflat_3+    \right.\\
	& \left. \qquad \qquad - \partial_{q_{1}}\partial_{q_{2}}\psiflat_{2R} \right] \label{introref}\,, \\
	\nonumber
	\begin{tikzpicture}
			\coordinate (v1) at (-2, 0);
			\coordinate (v2) at (0, 0);
			\coordinate (v3) at (2, 0);
			\draw[thick] (v1) -- (v2) --(v3);
			\draw[fill] (v1) circle (.5mm);
			\node at (-2,-0.3) {$  q_{1} $};
			\node at (-1,-0.3) {$  p_{1} $};
			\node at (0,-0.3) {$  q_{2} $};
			\node at (1,-0.3) {$  p_{2} $};
			\node at (2,-0.3) {$  q_{3} $};
			\draw[fill] (v2) circle (.5mm);
			\draw[fill] (v3) circle (.5mm);
			\node at (-1.7, .3) {$\phi$};
			\node at (-.4, .3) {$ \phi'$};
			\node at (.4, .3) {$\phi$};
			\node at (1.7, .3) {$\phi$};
		\end{tikzpicture} &=\frac{(2+q_1\partial_{q_1})}{p_2^2} \left[ (2+q_3\partial_{q_3})(-1-p_2\partial_{p_2}-q_3\partial_{q_3}) \psiflat_3  -  \partial_{q_{2}}\partial_{q_{3}}\psiflat_{2L} \right]\,,\\
\nonumber
	\begin{tikzpicture}
			\coordinate (v1) at (-2, 0);
			\coordinate (v2) at (0, 0);
			\coordinate (v3) at (2, 0);
			\draw[thick] (v1) -- (v2) --(v3);
			\draw[fill] (v1) circle (.5mm);
			\node at (-2,-0.3) {$  q_{1} $};
			\node at (-1,-0.3) {$  p_{1} $};
			\node at (0,-0.3) {$  q_{2} $};
			\node at (1,-0.3) {$  p_{2} $};
			\node at (2,-0.3) {$  q_{3} $};
			\draw[fill] (v2) circle (.5mm);
			\draw[fill] (v3) circle (.5mm);
			\node at (-1.7, .3) {$\phi$};
			\node at (-.4, .3) {$ \phi$};
			\node at (.4, .3) {$\phi$};
			\node at (1.7, .3) {$\phi$};
		\end{tikzpicture}&=\frac{1}{p_{1}^{2} p_2^2} \left[ (2+q_1\partial_{q_1})(2+q_3\partial_{q_3})(-1-q_1\partial_{q_1}-p_1\partial_{p_1})(-1-q_3\partial_{q_3}-p_2\partial_{p_2}) \psiflat_3- \right.\\
		&\nonumber\quad   (2+q_3\partial_{q_3})(-1-q_3\partial_{q_3}-p_2\partial_{p_2})  \partial_{q_{1}}\partial_{q_{2}}\psiflat_{2R}+\\
		&\nonumber\quad \left. - (2+q_1\partial_{q_1})(-1-q_1\partial_{q_1}-p_1\partial_{p_1})\partial_{q_{2}}\partial_{q_{3}}\psiflat_{2L}+\partial_{q_{1}}\partial_{q_{2}}^2\partial_{q_{3}}\psiflat_{1} \right]\,.\\
\end{align}
Notice that in each case there are $2^{E_{ \phi' \phi'}+E_{\phi\phi}}$ contributions, namely $  1 $, $  2 $ or $  4 $, as expected on general grounds. Here we left the arguments of the seed wavefunction $  \psiflat $ implicit, but they are recovered by the expressions below:
\begin{align}
\psiflat_{3}&=\psiflat_{3}(q_{1},q_{2},q_{3}; p_1, p_2) \,, &
\psiflat_{2R}&=\psiflat_{2}(q_{1}+q_{2},q_{3}; p_2) \,,\\
\psiflat_{2L}&=\psiflat_{2}(q_{1},q_{2}+q_{3}; p_1) \,,&
\psiflat_{1}&=\psiflat_{1}(q_{1}+q_{2}+q_{3}) \,.
\end{align}


\paragraph{Four-site example: the flux capacitor} As an example of a four-site chain with four vertices and three edges (bulk-to-bulk propagators), we study the flux capacitor with only $ \phi'^{n}$ interactions, so that all half-edges are of the $ \phi'$ type. Since all half-edges have one time derivative, all propagators are of the type $  G_{\phi'\phi'} $ and we don't need any re-routing, only summing over all possible contractions. Using the following kinematical variables
\begin{align}
\begin{tikzpicture}\label{fluxfig}
			\coordinate (v1) at (-2, 1);
			\coordinate (v2) at (2,1);
			\coordinate (v3) at (0,-2);
			\coordinate (v4) at (0,0);
			\draw[thick] (v1) -- (v4);
			\draw[thick] (v2) -- (v4);
			\draw[thick] (v3) -- (v4);			
			\draw[fill] (v1) circle (.5mm) node[anchor=south]{$  q_{1} $};
			\draw[fill] (v2) circle (.5mm) node[anchor=south]{$  q_{2}$};
			\draw[fill] (v3) circle (.5mm) node[anchor=north]{$  q_{3} $};
			\draw[fill] (v4) circle (.5mm) ;
			\node at (-.7, .7) {$p_{1}$};
			\node at (.8, .7) {$p_{2}$};
			\node at (0.3, -1) {$p_{3}$};
			\node at (0.3, -0.2) {$q_{4}$};			
\end{tikzpicture}
\end{align}
the result is
\begin{align}
\psint^{\text{flux capacitor}}&= p_{1}^{2}p_{2}^{2}p_{3}^{2}\psiflat_{4}(q_{1},q_{2},q_{3},q_{4})-p_{1}^{2}p_{2}^{2}\psiflat_{3}(q_{1},q_{4}+q_{3},q_{2}) + \text{2 perm's}  \label{fluxcap} \\
&\quad  +p_{1}^{2}\psiflat_{2}(q_{1},q_{2}+q_{3}+q_{4})+\text{2 perm's}-\psiflat_{1}(q_{1}+q_{2}+q_{3}+q_{4})\,. \nonumber
\end{align}

\paragraph{General $  \phi'^{n} $ theory} From the few examples above it is straightforward to see what $  \psint $ will be for a \textit{general} tree diagram with only $  \phi'^{n} $ interactions: 
\begin{align}\label{psintphiprime}
\psint^{\phi'^{n}}=\sum_{\text{collapsings}}^{2^{I}} \left[ (-1)^{n_{c}}\psiflat_{n-n_{c}} \prod_{m}^{I-n_{c}}p_{m}^{2}\right]\,.
\end{align}
Here the sum is over the $  2^{I} $ possible ways of collapsing any number $  n_{c} $ of the total $  I $ edges with a factor of $  (-1) $ for each of the collapsed edges and a factor of $  p_{m}^{2} $ for each of the $  I-n_{c} $ un-collapsed edges.


\section{Step 3: The final de Sitter wavefunction $\psifin$} \label{sec:5}

In the last step, Step 3, we specify the interactions that take place at each of the vertices and determine the final operators we apply to obtain the desired de Sitter wavefunction.  \\

The function $\psint$ is dictated by the skeleton and number of time-derivatives on half-edges.  From this point we can specify at each vertex an arbitrary $SO(3)$ invariant interaction which produces enough powers of conformal time to obey the bound in \eqref{bound}.  Because we start from the function $\psiflat$, which does not have the measure factor $  \sqrt{-g}=(\eta H)^{-4} $, it need be the case that each vertex carries four or more powers of conformal time. In the next section we will discuss a generalisation to couplings in general relativity (GR), some of which have negative powers of $ \eta$. Therefore, to get the correct result for a given interaction, we need to count the powers in excess of $  \eta^{4} $ and produce them with derivatives acting on external energies. This net power of conformal time $\eta_A$ at a given vertex is $d_A-4$, where $  d_{A} $ (see \eqref{bound}) is twice the number of time derivatives (since $  \dot K \sim \eta K' \sim \eta^{2} $) plus the number of spatial derivatives (since $  \partial_{i}\partial_{j}g^{ij}\sim \eta^{2} $). The minus four comes from the $  \sqrt{-g}\sim \eta^{-4} $ measure of the spacetime integral. Therefore, we need to generate $d_A-4$ additional powers of $  \eta_{A} $ with derivatives with respect to external kinematics\footnote{Since $  q_{A} $ is the sum of the external energies $  k_{a} $ for the legs attached to the vertex $  A $, we could alternatively take this derivative with respect to any of these $  k_{a} $.} $  q_{A} $. Furthermore, we also need to account for the fact that the bulk-to-boundary propagators for external legs are those in \eqref{masslessmodefct}, which differ from the flat space propagators in \eqref{Kflat} by a factor of $  (1-ik\eta) $. This missing factor can be simply generated by the acting with the differential operator $  (1-k_{b}\partial_{q_{A}}) $ where $  k_{b} $ is the energy of an external leg that is attached to a vertex $  A $ with total external energy $  q_{A} $.

Given the above discussion, we are now in the position of specifying the differential operator that need to act on each vertex $ A$:
\begin{equation}
	\label{etapower}
		\Delta_A = i^{4-d_A}\left( \prod_{a\in K'} k_{a}^{2} \right) \left( \prod_{b \in K} \left(1-k_b \partial_{q_A}\right) \right) \frac{\partial^{d_A-4}}{\partial q_A^{d_A-4}}\,.
\end{equation}
Here the first product goes over all external energies $  k_{a} $ of external legs that have one time derivative (bulk-to-boundary propagator $  K' $), while the second goes over all external energies $  k_{b} $ of external legs without any time derivatives (bulk-to-boundary propagator $  K $). An operator $  \Delta_{A} $ must be applied for each vertex $A$. Notice that we are assuming throughout that the external mode functions are all massless de Sitter mode functions as in \eqref{masslessmodefct}.  The counting of derivatives is based on the uncollapsed graph, as the intermediate wave-function accounted already for all effects of collapsing.

Additionally, we must account for all contractions of spatial momenta and possibly polarization tensors that are dictated by the nature of the interaction at each vertex. These are given by the contractions of spatial vectors dictated by the spatial derivatives at each vertex. We denote this additional factor for all vertices collectively by $F$. In this way, we can express the final de Sitter wavefunction coefficient $\psifin$ as 
\begin{align}\label{psisev}
	\psifin = F \left( \prod\limits_A \Delta_A \right)   \psint\,,
\end{align}
with the operator $  \Delta_{A} $ defined in \eqref{etapower} and $  \psint $ the output of Step 2.

 
\subsection{General properties}

The tree-level contributions to wavefunction coefficients possess general properties that become manifest in our differential representation. We highlight some of these properties below.


\paragraph{Scale invariance} Because of scale invariance, all contributions to the dS $ n$-point wavefunction coefficient $ \psi_{n}$ must\footnote{This scaling is violated by logarithmic corrections, which appear already at tree-level. The variation of logarithmic terms are analytical in two or more momenta (not energies) and therefore correspond to contact terms in position space. This mirrors the structure of Weyl anomalies in CFT's, as expected from AdS/CFT \cite{Henningson:1998gx}.} scale as $ k^{3}$. This can be seen in the differential representation as follows. First notice that for a diagram with $ V$ vertices
the flat spacetime seed function scales as 
\begin{align}
\psiflat_{V}\sim k^{1-2V}\,.
\end{align}
Then, it's easiest to account in one go for both Step 2 and the factor of $ k^{2}$ in Step 3 corresponding to a time derivative on an external leg. This can can be thought of as a $ k$ for every $ \phi'$ half-edge or $ \phi'$ external leg, a $ k^{-1}$ for every $ \phi$ half-edge, and a factor of $ k$ for every external leg. This correctly accounts for both the $ P$ factor in \eqref{Pfact} and for the two prefactors in \eqref{etapower}. Finally, the derivative in \eqref{etapower} gives a $ k^{4-d}$ for every vertex. Putting it all together we have
\begin{align}
\frac{\partial \ln \psi_{n}}{\partial \ln k}&\sim 1-2V+n+\sum_{A}^{V}(n_{\partial_{i}}+n_{\partial_{t}}-n_{\phi})+\sum_{B}^{V}(4-d_{B})\\
&\sim 1-2V+n+\sum_{A}^{V}(n_{\partial_{i}}+n_{\partial_{t}}-n_{\phi}+4-2n_{\partial_{t}}-n_{\partial_{i}})\\
&\sim 1-2V+n+4V-\sum_{A}n_{A}\\
&\sim 1+2V+n-(2I+n)=1+2(V-I)=3\,,
\end{align}
where $ n_{\phi}$ is the number of fields \textit{without time derivatives} in a given vertex $ A$, such that $ n_{A}=n_{\phi}+n_{\partial_{t}} $ is the valency of that vertex\footnote{To avoid the cumbersome notation $n_{\partial_{t}}^{A} $, $ n_{\phi}^{A}$ etc. we have left implicit the dependence on $ A$ of $ n_{\phi}$, $ n_{\partial_{t}}$ and $ n_{\partial_{i}}$.}. Also, in the fourth and fifth steps we used the graph identities
\begin{align}
\sum_{A}n_{A}&=2I+n\,, & V&=I+1\,.
\end{align}

\paragraph{Partial and total energy poles} The differential representation makes it clear that, by construction, the result is a rational function\footnote{Indeed the differential representation discussed so far is valid when all interactions obey $ d\geq 4$ in which case no logarithmic terms can arise.}. The only allowed poles are those that already exist in the seed wavefunction $ \psiflat$. The poles in $ \psiflat$ are constrained to be very specific linear combinations of energy variables determined by the following construction \cite{cosmopoly}. Consider a general ``skeleton'' diagram, namely a diagram where all external legs have been removed. As always we associate an energy $ q_{A}$ to each vertex and an energy $ p_{m}$ to each edge. Let $\gamma$ be a generic \textit{connected} sub-diagram and let $\partial \gamma$ be the set of all edges that are (\textit{i}) attached to vertices in $\gamma$ and (\textit{ii}) \textit{not} included in $\gamma$. The corresponding partial energy variable is defined by
\begin{align}
E_{\gamma}\equiv \sum_{\gamma }q_{A} +\sum_{m\in \partial \gamma}p_{m}\,.
\end{align}
The only possible poles in $ \psiflat_{V}$ and hence in the full $ \psi_{n}$ are partial energies of some connected sub-diagram. This includes as a limiting case the total energy $k_{T}= \sum q_{A}=\sum_{a} k_{a}$ corresponding to the sub-diagram in question being the full diagram itself. Despite appearances, the factor $ p_{m}^{-2}$ in \eqref{Pfact} does not introduce poles in the edge energies $ p_{m}$ because these are canceled by zeros in the associated bulk-to-bulk propagators. In fact, the absence of these divergences can be used as a starting point to derive the manifestly local test, as shown in \cite{MLT}.


\paragraph{Order of the total and partial energy poles} The order of the total and partial energy poles are fixed by the mass dimensions of the vertices involved in the corresponding diagram. In general, if
\begin{align}
\lim_{E_{\gamma}\to 0} \psi_{n} \sim \frac{1}{ E_{\gamma}^{\text{p}_{\gamma}}}
\end{align}  
we call\footnote{We use the non-italicised ``p'' to avoid confusion with the edge energy $ p$.} $ \text{p}_{\gamma} $ the \textit{order} of the corresponding partial energy pole. This obeys the upper bound
\begin{align}
\text{p}_{\gamma}\leq 1+ \sum_{A\in \gamma} \left(\dimop_{A}-4 \right)\,.
\end{align}
where $ \dimop_{A} $ is the mass dimension of the vertex $ A$, given by\footnote{Since $ n_{\phi}$ counts only the $ \phi$'s without time derivatives, we need a factors of $ 2$ in front of $ n_{\partial_{t}}$ to count the $ \phi$'s with one time derivative.} 
\begin{align}
\dimop_{A}= n_{\phi}+ 2 n_{\partial_{t}} +n_{\partial_{i}} \,.
\end{align}
When p corresponds to the total energy $k_{T} $, this upper bound on the order of the corresponding pole was derived in \cite{BBBB} using scale invariance and dimensional analysis. When p corresponds to the order of a partial energy poles, a bound can be derived using the cosmological cutting  rules \cite{Melville:2021lst}, which fix all terms with partial energy divergences \cite{MLT} using unitarity. Instead, here we derive this bound from our differential representation. The seed wavefunction has a simple pole in all partial energies. The order of this pole increases by one when it is hit by a differential operator in Step 2 and Step 3 of our construction. In Step 2 we need to apply the re-routing operator for every internal half-edge of the $ \phi$ type. This operator involves at most one derivative and therefore can raise the order of the pole by at most one. Notice that one can always choose the re-routing such that the half-edges external to a given sub-diagram never flows through that sub-diagram. Hence we need to only sum over the half-edges contained in a given sub-diagram. We can also increase the partial energy pole by one with the derivative term $ k_{b}\partial_{q_{A}}$ in \eqref{etapower} for every external leg of the $\phi $ type. Finally, the $ \partial_{q_{A}}$ operators in \eqref{etapower} can increase the order by $ d_{A}-4$. Putting it all together we find
\begin{align}
\text{p}_{\gamma}&\leq 1 +\sum_{A} n_{\phi}^{\text{internal}} + \sum_{B} n_{\phi}^{\text{external}} +\sum_{C} \left( d_{C}-4 \right) \\
&= 1+\sum_{A}\left(  n_{\phi}+n_{\partial_{t}} +n_{\partial_{1}}-4  \right)= 1+ \sum_{A\in \gamma} \left(\dimop_{A}-4 \right)\,,
\end{align}
which proves our claim.


\paragraph{Manifestly local test} A theory is said to be manifestly local when all interactions are products of a fields and and a finite number of their derivatives at the same spacetime point. The tree-level wavefunction coefficients for manifestly local theories of massless scalars and gravitons in dS must obey the simple condition (see \cite{MLT} for more details)
\begin{align}\label{mlt}
\partial_{k_{a}} \psi_{n}(k_{1},\dots,k_{a},\dots, k_{n}; \dots)\big|_{k_{a}=0}=0\,.
\end{align}
where in taking the derivative one should keep fixed all internal energies $ p_{m}$, all the other external energies $ k_{b}$ with $ b\neq a$ and all contractions of spatial momenta $ \v{k}_{a}\cdot \v{k}_{b}$. In other words, there is no linear term in $ k_{a}$ when Taylor expanding $ \psi_{n}$ around $ k_{a}=0$. It is straightforward to see that our differential representation indeed satisfies this manifestly local test. From the form of $ \Delta_{A}$ in \eqref{etapower} it is immediately clear that for all external legs with a time derivative $ \psi_{a} \sim \O(k_{a}^{2})$ as $ k_{a}\to 0$, which therefore satisfies \eqref{mlt}. For external legs without time derivatives we notice that
\begin{align}
\partial_{k_{a}} \left[ \prod_{b} (1-k_{b}\partial_{q_{A}}) F\right]_{k_{a}=0}&=\partial_{k_{a}} \left[ \prod_{b} (1-k_{b}\partial_{k_{b}}) F\right]_{k_{a}=0}\\
&=\left[ -\prod_{b\neq a}  (1-k_{b}\partial_{k_{b}}) \partial_{k_{a}} F+ \prod_{b} (1-k_{b}\partial_{k_{b}}) \partial_{k_{a}} F\right]_{k_{a}=0}=0\,.
\end{align}


\subsection{Examples}

To clarify our whole procedure we provide here the differential representations for a series of wavefunction coefficients.

\paragraph{One-site chain: contact diagrams} As a natural starting point, we now compute contact diagrams. Even though these are quite easily computed also in the time-integral representation, they are simple enough to provide a good example of our procedure. For example consider the contact contribution to $  \psi_{n} $ from the interaction 
\begin{align}\label{phipn}
\L_{\text{int}}=\frac{g}{n!}a^{4-n}(\phi')^{n}\,,
\end{align}  
which has mass dimension $  \dimop = 2n $. The flat space seed wavefunction of Step 1 is simply $  \psiflat_{1}=1/q $ where $  q $ is the sum of all external energy and so can also be written as the more familiar total energy, $  q=k_{T} $. Since there are no propagators, there is no need to perform Step 2. Finally, in Step 3 we hit $  \psiflat $ with the operator in \eqref{etapower} where $  d=2n $:
\begin{align}
\psi_{n}&=gi^{2n-4}\left( \prod_{a}^{n}k_{a} \right)^{2}\partial_{k_{T}}^{2n-4} \frac{1}{k_{T}}\\
&=gi^{2n-4}(2n-4)!\,\frac{\left( \prod_{a}^{n}k_{a} \right)^{2}}{k_{T}^{2n-3}}\,.\label{agree}
\end{align}
We remind the reader that we work in units of the Hubble parameter, which is constant in de Sitter, so $  H=1 $. Notice that the order $  p $ of the total energy pole $  1/k_{T}^{p} $ matches what's expected on the general grounds of scale invariance and dimensional analysis \cite{BBBB}
\begin{align}\label{gen}
p=1+\sum_{A} \left( \dimop_{A}-4 \right)\,, 
\end{align}
where in general the sum goes over all vertices $  A $ in the diagram. For the contact interaction in \eqref{phipn} this correctly reproduces $  p=2n-3 $, in agreement with \eqref{agree}. \\

As another example consider 
\begin{align}
\L_{\text{int}}=g a^{4-2n}\,(\partial_{i}\phi \partial_{i}\phi)^{n}\,,
\end{align}  
which has mass dimension $  \dimop=4n $. The seed function of Step 1 is again $  1/k_{T} $ and Step 2 is trivial. Finally, Step 3 gives us the final answer
\begin{align}
\psi_{2n}=F i^{2n-4}\left[ \prod_{a=1}^{2n}\left(  1-k_{a}\partial_{k_{T}}\right) \right] \partial_{k_{T}}^{2n-4}\frac{1}{k_{T}}\,,
\end{align}
where $  F $  is the sum over all permutations $  \sigma $ of the $  2n $ external legs of the product of all possible pairings 
\begin{align}
F=\sum_{\sigma}\left( \v{k}_{\sigma_{1}}\cdot \v{k}_{\sigma_{2}} \right) \left( \v{k}_{\sigma_{3}}\cdot \v{k}_{\sigma_{4}}  \right)\dots\left(  \v{k}_{\sigma_{2n-1}}\cdot \v{k}_{\sigma_{2n}} \right)\,.
\end{align}
The expression for $  \psi_{2n} $ can be re-written in terms of elementary symmetric polynomials $  e_{a} $ for $  2n $ variables
\begin{align}
e_{0}&=1\,,\\
e_{1}&=\sum_{1\leq j_{1}\leq 2n} k_{j_{1}}=k_{T}\,,\\
e_{2}&=\sum_{1\leq j_{1}<j_{2}\leq 2n} k_{j_{1}}k_{j_{2}}\,,\\
\dots& \nonumber \\
e_{a}&=\sum_{1\leq j_{1}<j_{2}<\dots<j_{a}\leq2n} k_{j_{1}}k_{j_{2}}\dots k_{j_{a}}\,,\\
e_{2n}&=k_{1}k_{2}\dots k_{2n}\,.
\end{align}
The final wavefunction coefficient then becomes
\begin{align}
\psi_{2n}=F i^{2n-4}\sum_{a=0}^{2n}e_{a}\frac{(2n+a-4)!}{k_{T}^{2n+a-3}}\,,
\end{align}
for which the largest total energy pole $  p=4n-3 $ is in agreement with the general result in \eqref{gen}.


\begin{figure}
	\begin{center}
		\begin{tikzpicture}
			\coordinate (v1) at (-3, 0);
			\coordinate (v2) at (3, 0);
			\coordinate (b4) at (4, 2);
			\coordinate (b3) at (2, 2);
			\coordinate (b2) at (-2, 2);
			\coordinate (b1) at (-4, 2);
			\coordinate (vp) at (0, -.5);
			\draw[thick] (b1) -- (v1) -- (b2);
			\draw[thick] (b3)  --  (v2) -- (b4);
			\draw[thick] (v1) -- (v2);
			\draw[very thick] (-6, 2) -- (6, 2);
			\node at (0, -0.5) {$p $};
			\node at (-4, 2.5) {$\v{k}_{1}$};
			\node at (-2, 2.5) {$\v{k}_{2}$};
			\node at (2, 2.5) {$\v{k}_{3}$};
			\node at (4, 2.5) {$\v{k}_{4}$};
			\node at (-3.5,0) {$  q_{1} $};
			\node at (3.5,0) {$  q_{2} $};
		\end{tikzpicture}
	\end{center}
	\caption{The scalar quartic wavefunction $  \psi_{4}^{s} $ from the exchange of the same scalar in the $  s $-channel.}
	\label{psi4s}
\end{figure}
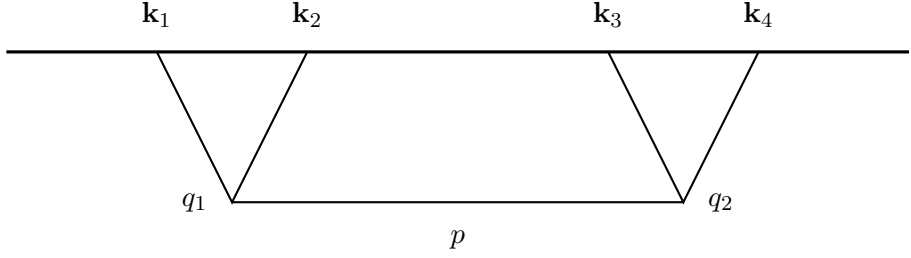

\paragraph{Two-site chain: single-exchange diagram} A more potent example of our differential representation is the calculation of the most important wavefunction coefficients generated by the leading operators in the Effective Field Theory of inflation \cite{Cheung:2007st}. Here we will discuss the quartic exchange wavefunction $  \psi_{4}^{s} $ in the $  s $-channel. Other channels are simply obtained by permutations. We will consider all the possible combinations of the cubic, shift-symmetric interactions 
\begin{align}
\L_{int} = \frac{g_{1}}{3!} a (\phi')^{3} + \frac{g_{2}}{2}a  \phi' (\partial_{i }\phi)^{2} \,.
\end{align}
These are expected to be the leading ones in the decoupling limit. Let's start with the exchange generated by two insertions of $  (\phi')^{3} $. In Step 1 we write down the corresponding flat space wavefunction
\begin{align}\label{flatswves}
\psiflat_{2}&=\frac{1}{(q_{1}+q_{2})(q_{1}+p)(q_{2}+p)}\,, &\psiflat_{1}&=\frac{1}{q_{1}+q_{2}}\,,
\end{align}
where the single vertex wavefunction $  \psi_{\text{flat},1} $ will be needed in the contractions of the $  \ex{\phi'\phi'} $ propagator. As depicted in Figure \ref{psi4s}, we have two external legs attached to each vertex, so we have the relations 
\begin{align}
q_{1}&=k_{1}+k_{2}\,,&  q_{2}&=k_{3}+k_{4}\,,&  p&=|\v{k}_{1}+\v{k}_{2}|\,.
\end{align}
Moving on to Step 2, we can use the result of \eqref{phipphip} to write 
\begin{align}\label{psint2}
\psint^{\phi'\phi'}&= p^{2} \psiflat_{2} - \psiflat_{1} \\
&=    \frac{p^{2}}{(q_{1}+q_{2})(q_{1}+p)(q_{2}+p)}-\frac{1}{q_{1}+q_{2}}\,.
\end{align}
Finally, moving on to Step 3, we have to apply one differential operator $  \Delta_{1} $ to the first vertex with the power of conformal time given by $  d_{1}=2\times 3+0=6 $ and another $  \Delta_{2} $ to the other vertex with $  d_{2}=6 $. Hence, we find our final differential representation for this wavefunction coefficient: 
\begin{align}
\psi_{4}^{g_{1}g_{1}}=g_{1}^{2}(k_{1}^{2}k_{2}^{2}k_{3}^{2}k_{4}^{2})\partial_{q_{1}}^{2}\partial_{q_{2}}^{2}\psint^{\phi'\phi'}\,.
\end{align}
This is a very compact representation of the result and the derivatives can be calculated in a software such as Mathematica in a fraction of a second. The result is conveniently written in terms of $  k_{T}=\sum_{a} k_{a} $, $  E_{L}=q_{1}+p $ and $  E_{R}=q_{2}+p $ in the form
\begin{align}
\psi_{4}^{g_{1}g_{1}}&=4g_1^2\left( k_{1}k_{2}k_{3}k_{4}p \right)^{2}\left[  \frac{6}{k_{T}^{5}E_{L}E_{R}}+ \frac{3}{k_{T}^{4}E_{L}E_{R}}\left(  \frac{1}{E_{L}}+\frac{1}{E_{R}}\right)+\frac{1}{k_{T}^{3}E_{L}E_{R}}\left(  \frac{1}{E_{L}}+\frac{1}{E_{R}}\right)^{2}+\right. \nonumber\\
& \hspace{3cm} \left. +\frac{1}{k_{T}^{2}E_{L}^{2}E_{R}^{2}}\left(  \frac{1}{E_{L}}+\frac{1}{E_{R}}\right)+\frac{1}{k_{T}E_{L}^{3}E_{R}^{3}}-\frac{6}{p^{2}k_{T}^{5}}\right]\,.
\end{align}
As a second case we consider inserting the $  (\phi')^{3} $ interaction on the left vertex, corresponding to $  q_{1} $ and $  \phi' (\partial_{i}\phi)^{2} $ on the other vertex, corresponding to $  q_{2} $. Step 1 is just the same as in the previous case and gives the flat space wavefunctions in \eqref{flatswves}. In Step 2 we encounter two types of propagators, the one corresponding to $  \ex{\phi'\phi'} $, which we computed in \eqref{psint2}, and the one corresponding to $  \ex{\phi'\phi} $, which arises when the time derivative in $  \phi' (\partial_{i}\phi)^{2} $ hits an external leg. For this latter case we can use \eqref{phipphidraw}
\begin{align}
\psint^{\phi'\phi}&=(2+q_2\partial_{q_2}) \psiflat_{2}\\
&=\frac{   q_{1}q_{2}+p\left( 2q_{1}+q_{2} \right) }{(p+q_{1})(p+q_{2})^{2}(q_{1}+q_{2})^{2}}\,.
\end{align}
We are finally ready to perform Step 3 by acting with the differential operator in \eqref{etapower} on the two vertices, which have respectively $  d_{1}=2\times 3+0=6 $ and $  d_{2}=2\times 1+2=4 $. Accounting for the two possible propagators and the factor of $  F $ from the contraction of spatial derivatives on the second vertex, we find
\begin{align}
\psi_{4}^{g_1 g_2}&= -g_1 g_2 k_{1}^{2}k_{2}^{2}\partial_{q_{1}}^{2}\left\{  \v{k}_{3}\cdot \v{k}_{4}(1-k_{3}\partial_{q_{2}})(1-k_{4}\partial_{q_{2}})\psint^{\phi'\phi'}+ \right. \nonumber \\
&\left. \quad +\left[  \v{k}_{4}\cdot \v{p}\, k_{3}^{2}\left(  1-k_{4}\partial_{q_{2}}\right)+ \v{k}_{3}\cdot \v{p}\,k_{4}^{2}\left(  1-k_{3}\partial_{q_{2}}\right)\right] \psint^{\phi'\phi}\right\}\,
\end{align}
As our last example, we derive the differential representation for two insertions of $  \phi' (\partial_{i}\phi)^{2} $. This is the longest expression because it involves also a third type of exchange propagator without any time derivatives. The intermediate wavefunction for that case was worked out in Section \ref{step2ex}.
Following the prescribed rules, the full wavefunction is
\begin{multline}
	\psi_4^{g_2 g_2} = g_2^2\big [(\v{k}_1 \cdot \v{k}_2 )(\v{k}_3 \cdot \v{k}_4) (1-k_1\partial_{q_1})(1-k_2\partial_{q_1})(1-k_3\partial_{q_2})(1-k_4\partial_{q_2})\psint^{\phi'\phi'}\\+k_3^2(\v{k}_1 \cdot \v{k}_2 )(\v{p} \cdot \v{k}_4)(1-k_1\partial_{q_1})(1-k_2\partial_{q_1})(1-k_4\partial_{q_2})\psint^{\phi'\phi}\\
	+k_2^2(\v{k}_1 \cdot \v{p})(\v{k}_3 \cdot \v{k}_4)(1-k_1\partial_{q_1})(1-k_3\partial_{q_2})(1-k_4\partial_{q_2})\psint^{\phi\phi'}+\\
	 k_1^2k_3^2(\v{k}_2 \cdot \v{p})(\v{k}_4 \cdot \v{p})(1-k_2 \partial_{q_1})(1-k_4\partial_{q_2})\psint^{\phi\phi}\big]+ \text{3 perms},
\end{multline}
where it's understood that in each exchange class we sum over the necessary permutations of labels for external lines and the three $  \psint $ are given in \eqref{phipphip}, \eqref{phipphidraw} and \eqref{phiphidraw}. Here the three permutations correspond the exchange of $  1 \leftrightarrow 2 $, $  3\leftrightarrow 4 $ and the combination of these two.


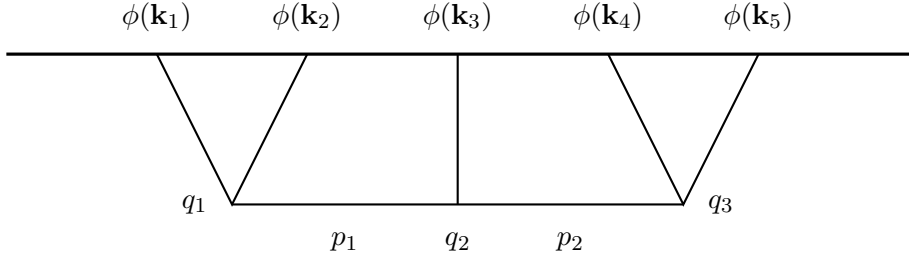
\begin{figure}
	\begin{center}
		\begin{tikzpicture}
			\coordinate (v1) at (-3, 0);
			\coordinate (v2) at (0, 0);
			\coordinate (v3) at (3, 0);
			\coordinate (b4) at (4, 2);
			\coordinate (b3) at (2, 2);
			\coordinate (b2) at (-2, 2);
			\coordinate (b1) at (-4, 2);
			\coordinate (b5) at (0,2);
			\coordinate (vp) at (0, -.5);
			\draw[thick] (b1) -- (v1) -- (b2);
			\draw[thick] (b3)  --  (v3) -- (b4);
			\draw[thick] (v1) -- (v2) -- (v3);
			\draw[thick] (v2) -- (b5);
			\draw[very thick] (-6, 2) -- (6, 2);
			\node at (-1.5, -0.5) {$p_{1} $};
			\node at (1.5, -0.5) {$p_{2} $};
			\node at (-4, 2.5) {$\phi(\v{k}_{1})$};
			\node at (-2, 2.5) {$\phi(\v{k}_{2})$};
			\node at (0, 2.5) {$\phi(\v{k}_{3})$};
			\node at (2, 2.5) {$\phi(\v{k}_{4})$};
			\node at (4, 2.5) {$\phi(\v{k}_{5})$};
			\node at (-3.5,0) {$  q_{1} $};
			\node at (3.5,0) {$  q_{3} $};
			\node at (0,-0.5) {$  q_{2} $};
		\end{tikzpicture}
	\end{center}
	\caption{The scalar quintic wavefunction $  \psi_{5}^{s} $ from the double exchange of the same scalar. 	\label{fig3}}
\end{figure}

\paragraph{Three-site chain: double-exchange diagram} The previous examples discussed diagrams for which the integral bulk representation is still a practical computational tool because there are at most two nested time integrals. Here we want to show that the differential representation remains quite simple and easy to derive also when the number of nested time integrals grows, which makes a brute force bulk time integration very slow. As an example, we discuss the (tree-level) five-point function from three insertions of $  (\phi')^{3} $. For Step 1 we need the flat space wavefunction corresponding to the three site chain, which we computed in \eqref{psiflat3} (see Figure \ref{fig3} for the definition of kinematical variables)
\begin{align}
\psiflat_{3}&=\frac{1}{(q_1+q_2+q_3)(q_1+p_1)(q_2+p_1+p_2)(q_3+p_2)}\left(   \frac{1}{q_1+q_2+p_2} +  \frac{1}{q_2+q_3+p_1}  \right)\,.
\end{align}
The corresponding intermediate wavefunction was computed in \eqref{phip4},
\begin{align}
	\psi_{\text{int}}^{\phi'\phi'\phi'\phi'}  = p_1^2p_2^2\psi_3^{\text{flat}}-p_1^2\psi_{2L}^{\text{flat}}-p_2^2\psi_{2R}^{\text{flat}}+\psi_1^{\text{flat}}\,,
\end{align}
where $\psi_{2L}^{\text{flat}}$ is the flat space wave-function for a two-site chain for thediagram made from collapsing the $p_2$ exchange edge and $\psi_{2R}^{\text{flat}}$ is that made from collapsing the $p_1$ exchange edge.  Since the $  \phi'^{3} $ interaction has $  d=6 $, the desired de Sitter wavefunction is simply
\begin{align}
\psi_{5}^{g_{1}^{3}} = - \left(  \prod_{a=1}^{5}k_{a}^{2} \right)\partial_{q_1}^2\partial_{q_2}^2\partial_{q_3}^2 \psi_{\text{int}}^{\phi'\phi'\phi'\phi'} \,.
\end{align}

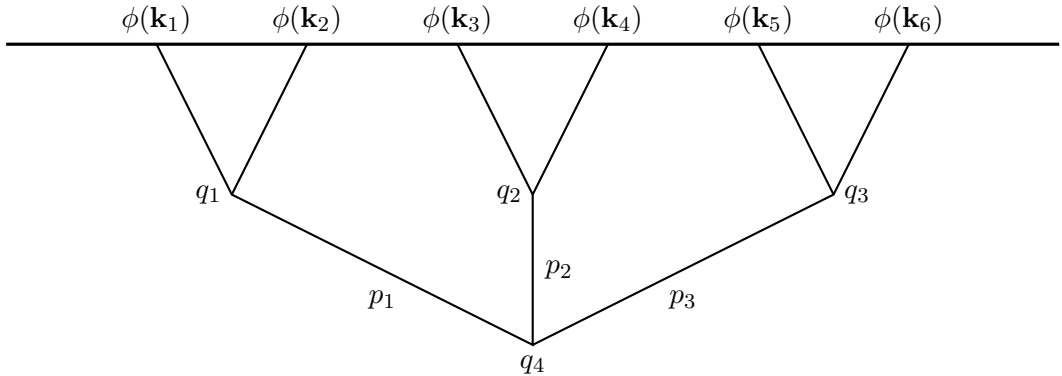
\begin{figure}
	\begin{center}
		\begin{tikzpicture}
			\coordinate (v1) at (-4, -2);
			\coordinate (v2) at (0, -2);
			\coordinate (v3) at (4, -2);
			\coordinate (v4) at (0, -4);
			\coordinate (b1) at (-5, 0);
			\coordinate (b2) at (-3, 0);
			\coordinate (b3) at (-1, 0);
			\coordinate (b6) at (5, 0);
			\coordinate (b5) at (3, 0);
			\coordinate (b4) at (1, 0);
			\draw[thick] (b1) -- (v1) node[anchor=east]{$  q_{1} $} -- (v4) -- (v2) -- (b3);
			\draw[thick] (b2) -- (v1);
			\draw[thick] (b4) -- (v2) node[anchor=east]{$  q_{2} $};
			\draw[thick] (b5) -- (v3) node[anchor=west]{$  q_{3} $};
			\draw[thick] (b6) -- (v3) -- (v4) node[anchor=north]{$  q_{4} $};
			\draw[very thick] (-7, 0) -- (7, 0);
			\node at (b1) [anchor=south]{$  \phi(\v{k}_{1}) $};
			\node at (b2) [anchor=south]{$  \phi(\v{k}_{2}) $};
			\node at (b3) [anchor=south]{$  \phi(\v{k}_{3}) $};
			\node at (b4) [anchor=south]{$  \phi(\v{k}_{4}) $};
			\node at (b5) [anchor=south]{$  \phi(\v{k}_{5}) $};
			\node at (b6) [anchor=south]{$  \phi(\v{k}_{6}) $};
			\node at (-2, -3.4) {$p_1$};
			\node at (2, -3.4) {$p_3$};
			\node at (0.35, -3) {$p_2$};
		\end{tikzpicture}
	\end{center}
	\caption{The flux capacitor contribution to the 6-point function $  \psi_{6} $ from the interaction $  \phi'^{3}/3! $ \label{fluxfig2}}
\end{figure}

\paragraph{Four-site chain: the flux capacitor} Using the result for the flux-capacitor intermediate wavefunction $ \psint$ in \eqref{fluxcap}, and assuming that all interactions are of the form $ \phi'^{3}/3!$, it is immediate to write down the six-point wavefunction coefficient associated with the diagram in Figure \ref{fluxfig2}
\begin{align}
\psi_{6}=\prod_{a=1}^{6} k_{a}^{2}   \prod_{A=1}^{4}\partial_{q_{A}}^{2}\,\psint\,.
\end{align}

\paragraph{General $  \phi'^{n} $ theory} For a general theory with $  \phi'^{n} $ interactions for any set of $  n $'s, we computed the intermediate wavefunction in \eqref{psintphiprime} for any tree-level diagram with an arbitrary number of internal and external propagators. Using that fact that a $  \phi'^{n} $ interaction has $  d=2n $, we find that the final dS wavefunction is simply given by 
\begin{align}
\psi_{n}=\prod_{a=1}^{n} k_{a}^{2}   \prod_{A=1}^{V}(-1)^{n_A}\partial_{q_{A}}^{2n_{A}-4}\,\psint\,,
\end{align}
where $  n_{A} $ is the valency of the vertex $  A $ and $  \psint $ was given in \eqref{psintphiprime}.


\section{Minimal coupling to gravity}\label{sec:minimal}

In this section, we show how to generalise our differential representation to describe gravitational interactions, which contain terms with only two spatial derivatives and hence violate the inequality in \eqref{bound}. This issue can be overcome both for contact and exchange diagrams at tree level. As concrete examples, we discuss the quartic scalar and quartic graviton wavefunction coefficients induced by graviton exchange in GR. 


\subsection{Contact diagrams}
To describe minimal coupling to gravity, we need to accommodate interactions with two derivatives. Two time derivatives are already accounted for by our previous treatment because they satisfy \eqref{bound}. Assuming parity, the only other option\footnote{When solving for the constrained lapse and shift one also generates non-manifestly local interactions for the graviton and for gravitationally coupled scalars. Counting an inverse Laplacian as contributing $ -2$ to $ n_{\partial_{i}}$, we have checked that these interactions all have $ d\geq 4$ and so fall within the regime of applicability of the techniques we have presented in the first part of the paper, where manifest locality was not assumed.} is an interaction with two spatial derivatives. Since this has $ d=2$ it violates \eqref{bound} and has the following time-dependent integrand (up to overall factors of $ \v{k}$)
\begin{equation}
	\label{minimalcoupling}
	i\int d\eta a^2 K^n\,,
\end{equation}
with $K$ the massless propagators in \eqref{masslessmodefct}. At the contact level, this interaction does not yield a finite wavefunction as it has a divergent imaginary part. This can be seen by counting the powers of $  \eta $. However, notice that, for parity-even interactions, it is only the real part of $  \psi $ that contributes to the correlator. This is IR finite despite the $  1/\eta $ divergence in the imaginary part of $  \psi $. Explicitly, the $n$-point contact diagram produces the integral
\begin{equation}
	\psi_c = i\int \frac{d\eta}{\eta^2}\prod\limits_{a = 1}^n (1-ik_a \eta)e^{ik_a \eta}\,.
\end{equation}
How can we obtain this $  1/\eta^{2} $ term by acting with derivatives on the flat-space seed wavefunction, where no such factor is present? We observe that the above can be re-expressed as
\begin{equation}
\label{minimalderiv}
	\psi_c = i\int d\eta \left[-\partial_\eta \left(\frac{e^{ik_T\eta}}{\eta}  \right)-\sum^{n}\limits_{p = 2} (-i \eta)^{p-2} e_p(\{k_a \})e^{ik_T\eta} \right]\,,
\end{equation}
where $e_p(\{k_a \})$ is the elementary symmetric polynomial of order $p$ in the $k_a$'s, e.g. $e_2(u, v, w) = uv+uw+vw$. 
The $\eta \to 0$ limit produces a divergence in the first term. If we consider the real part, the divergence drops out and we find
\begin{equation}
	\text{Re }\psi_c = -\frac{i}{2} \lim_{\eta_0 \to 0} \frac{e^{ik_T \eta_0}-e^{-ik_T\eta_0}}{\eta_0} -\sum\limits_{p = 2}^n e_p(\{ k_a\}) \frac{(p-2)!}{k_T^{p-1}}\,,
\end{equation} 
and so we have 
\begin{equation}
	\text{Re }\psi_c = k_T -\sum\limits_{p = 2}^n e_p(\{ k_a\}) \frac{(p-2)!}{k_T^{p-1}}\,.
\end{equation}
This is in agreement with the classic result in \cite{Maldacena:2002vr}.
For more general diagrams we need to accommodate exchange diagrams containing vertices with $ d=2$, of the same form as \ref{minimalcoupling}.  The total derivative leveraged in equation \eqref{minimalderiv} allows us to do this.


\subsection{Exchange Diagrams}

In an exchange diagram, the vertex in \eqref{minimalcoupling}, which has $ d=2$, will generically have $l$ bulk-to-boundary propagators and $n-l$ bulk-to-bulk propagators attached to it:
\begin{equation}
	\int \dots \int \frac{d\eta}{\eta^2}\prod\limits_{a = 1}^l (1-ik_a \eta)e^{ik_a \eta}\prod\limits_{m = 1}^{n-l}G(\eta, \eta_i', p_m)\,.
\end{equation}
As in the contact case, we can express this as 
\begin{equation}
	\int \dots\int d\eta\left[-\partial_\eta \left(\frac{e^{iq_A\eta}}{\eta}  \right)-\sum\limits_{p = 2}^l(-i\eta)^{p-2}e_p(\{k_a \})e^{iq_A\eta} \right]
\prod\limits_{m = 1}^{n-l}G(\eta, \eta_i', p_m)\,,
\end{equation}
where $  q_{A} = \sum^{l} k_{a} $. We can integrate by parts and observe that
\begin{equation}
	\lim_{\eta \to 0} \frac{1}{\eta}\prod\limits_{i = 1}^{n-l}G(\eta, \eta_i', p_m) = 0\,,
\end{equation}
which ensures the vanishing of the boundary term.  Therefore the portion of the time-integral from this vertex is 
\begin{equation}
\label{replacement}
	\int \dots\int d\eta e^{iq_A\eta}\left[\frac{1}{\eta} \partial_\eta-\sum\limits_{p = 2}^l(-i\eta)^{p-2}e_p(\{k_a \}) \right]
\prod\limits_{m = 1}^{n-l}G(\eta, \eta', p_m)\,.
\end{equation}
Applying this argument to diagrams with multiple vertices with $  d=2 $ we end up with
\begin{equation}
	\label{fullibp}
	\int \prod\limits_A d\eta_A e^{iq_A \eta_A}\left[\frac{1}{\eta_A} \partial_{\eta_A}-\sum\limits_{p = 2}^l(-i\eta_A)^{p-2}e_p(\{k_a \})  \right]\prod\limits_{m = 1}^IG(p_m)\,.
\end{equation}
Thus far we have merely performed the same integration by parts employed in the contact case at all vertices in an exchange graph.  In \eqref{fullibp} this is expressed as an operator in conformal time acting on $G=G_{\phi\phi}$ propagators.  That is, in the language of previous sections, we have time derivatives or powers of conformal time acting on half-edges of the $\phi$ type.  Notice that when $\partial_\eta$ acts on the propagator we get an additional factor of conformal time as in \eqref{propspp}, so the expression in \eqref{fullibp} manifestly satisfies the condition in \eqref{bound} at the vertex in question.   Indeed we see that the application of the derivative expands the time integral in $\eqref{replacement}$ into a sum over $n-l+1$ time integrals of the type computed in previous sections.\\

Generically, when the number of symmetric polynomial terms $l-p$ is greater than 1, we cannot write a ``local'' differential operator acting on the seed function (conversely this is possible for the leading $n = 3$ case which we consider below).   The difference between these is the $\Delta_{A}$ operator, which accounts for the overall power of conformal time at the vertex $A$. However, these differing operators will not in general commute with the differential operators associated with other parts of the diagram and therefore cannot be locally lumped with the operator coming from $ \frac{1}{\eta}\partial_\eta$.  Therefore, for general $n$ one must consider a sum of  $2^{V_g}$ diagrams where $V_g$ is the number of vertices with $  d=2 $. At each vertex we must apply the local expansion depicted in Fig. \ref{gravexpansion}. The expansion has two terms: one for the time derivative portion of \eqref{replacement} and one for the second term.   We can prescribe the associated differential operators for each of these.  For the first diagram, coming from the time derivative term, we have differential operators:
\begin{align}
\begin{cases}\label{spec1}
	\Delta_{\text{grav}}^{\phi'} =  \sum\limits_{m=1}^{n-l} p_m \prod\limits_{m' \neq m}\frac{1}{p_{m'}} \Delta_{\text{rr}}^{m'A}\,, \\
	\Delta_A = 1\,,
\end{cases}
\end{align}
where the sum runs over all $n-l$ bulk-to-bulk propagators that are hit by the time derivative, and the product is over the remaining propagators.  The second term has differential operators
\begin{align}
\begin{cases}\label{spec2}
	\Delta_{\text{grav}}^\phi =  \sum\limits_{p = 2}e_p(\{ k_a \}) \prod\limits_{m}\frac{1}{p_{m}} \Delta_{\text{rr}}^{mA}\,, \\
	\Delta_A = -\sum\limits_{p = 2} (-\partial_{q_A})^{p-2}\,.
\end{cases}
\end{align}
In these expression we have specified where the re-routing operator $  \Delta_{\text{rr}} $ starts from, namely from the half-edge $  mA $ which is the part of the edge $  m $ that ends on the vertex $  A $. These operators completely prescribe the computation of the uncollapsed contribution, but we still have to specify the collapsed contribution. \\

 The complete prescription requires specification of the complementary half-edges on each propagator. However, while remaining agnostic about $A'$ we can restate the relevant form of equation \ref{collapsedeq} for a vertex $A$:
 \begin{align}
 \label{collapsedgrav}
 	\text{collapsing } \phi\phi &\to \hspace{3mm} \left(\sum\limits_p e_p(\{ k_a\}) (-\partial_{q_{A}})^{p-2}\right)\eta_{A'}^{d_{A'
	}-4}\left[\frac{1}{p^2} \eta_A\eta_{A'}\delta(\eta_A-\eta_{A'})\right]\,,\\ 
	\text{collapsing } \phi'\phi' &\to  \hspace{3mm} \eta_{A'}^{d_{A'}-4}\left[-\delta(\eta_A-\eta_{A'})\right]\,.
 \end{align}

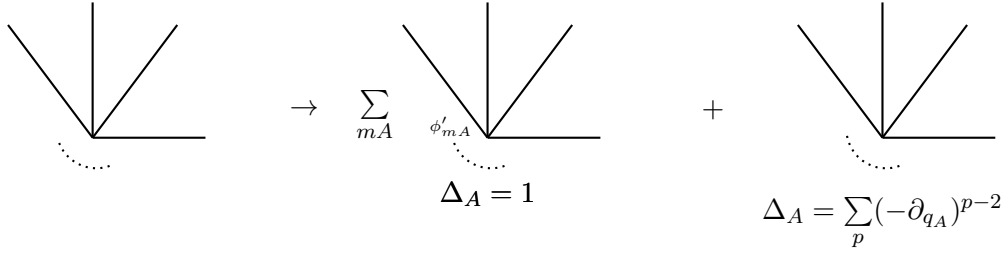
\begin{figure}
\label{gravexpansion}
	\begin{center}
		\begin{tikzpicture}[scale=.75]
			\coordinate (vL) at (-10, 0);
			\coordinate (oL1) at (-11.5, 2);
			\coordinate (oL2) at (-10, 2.4);
			\coordinate (oL3) at (-8.5, 2);
			\coordinate (oL4) at (-8, 0);
			\draw[thick] (vL) -- (oL1);
			\draw[thick] (vL) -- (oL2);
			\draw[thick] (vL) -- (oL3);
			\draw[thick] (vL) -- (oL4);
			\draw[thick, dotted] (-9.7,-.5) arc[x radius=.7, y radius=.7, start angle=290, end angle=200];
			\node at (-6.25, .5) {$\rightarrow$};
			\node at (-5, .4) {$\sum\limits_{mA}$};
			\coordinate (vM) at (-3,0);
			\coordinate (oM1) at (-4.5, 2);
			\coordinate (oM2) at (-3, 2.4);
			\coordinate (oM3) at (-1.5, 2);
			\coordinate (oM4) at (-1, 0);
			\draw[thick] (vM) -- (oM1);
			\draw[thick] (vM) -- (oM2);
			\draw[thick] (vM) -- (oM3);
			\draw[thick] (vM) -- (oM4);
			\draw[thick, dotted] (-2.7,-.5) arc[x radius=.7, y radius=.7, start angle=290, end angle=200];
			\node at (-3.65, .2) {\tiny$\phi'_{mA}$};
			\node at (-3, -1) {$\Delta_A = 1$};
			\coordinate (vM) at (4, 0);
			\coordinate (oM1) at (2.5, 2);
			\coordinate (oM2) at (4, 2.4);
			\coordinate (oM3) at (5.5, 2);
			\coordinate (oM4) at (6, 0);
			\draw[thick] (vM) -- (oM1);
			\draw[thick] (vM) -- (oM2);
			\draw[thick] (vM) -- (oM3);
			\draw[thick] (vM) -- (oM4);
			\draw[thick, dotted] (4.3,-.5) arc[x radius=.7, y radius=.7, start angle=290, end angle=190];
			\node at (-3, -1) {$\Delta_A = 1$};
			\node at (1, .5) {$+$};
			\node at (4, -1.5) {$\Delta_A = \sum\limits_p (-\partial_{q_A})^{p-2}$};
		\end{tikzpicture}
	\end{center}
	
	\caption{Expansion for vertices of type $\frac{1}{\eta^2}\prod (\phi_{ml})_i $ where  $\phi_{ml}$ denotes massless mode function.  The first term in the expansion denotes a sum over half-edges $Am$, namely the part of edge $ m$ ending on the vertex $A$, where that half-edge becomes a $\phi'$.  In the second term all half-edges are of the $ \phi$-type, corresponding to the ordinary massless mode functions.}
	\label{gravexpansion}
\end{figure}


\paragraph{n = 3} We now restrict to the case of $n = 3$ and provide some examples.  This involves the by now familiar collapsing procedure, coming from the $\phi'$ or $\phi$ that has been generated via the respective term in the integration by part being paired with $\phi$ or $\phi'$ at $A'$.  The leading gravitational interactions have $n = 3$. After integration by part, the vertex  is  
\begin{equation}
\label{replacementn3}
	\int \dots\int d\eta e^{iq_A\eta}\left[\frac{1}{\eta} \partial_\eta-e_2(\{k_a \})\delta_{l1} \right]
\prod\limits_{i = 1}^{l}G(\eta, \eta_i', p_m)
\end{equation}
where the Kronecker $\delta_{l1}$ manifests that we only get the second term inside the square brackets if we have precisely one exchange propagator attached to the vertex in question.  This case is particularly simple because of the absence of overall powers of $\eta$ at this vertex that need be accounted for and we can prescribe the differential operator locally, without the generic expansion over multiple diagrams described above.  In particular, the differential operator associated to a vertex is
\begin{equation}
	\label{deltagrav}
	\Delta_{\text{grav}}^A = \left[ \sum\limits_{m=1}^l p_m \prod\limits_{m' \neq m}\frac{1}{p_{m'}} \Delta_{\text{rr}}^{m'A} \right]-e_2(\{k_a \})\delta_{l1}\prod\limits_{m}^l \frac{1}{p_{m}} \Delta_{\text{rr}}^{mA}\,,
\end{equation}
where the sum and products over $ m$ and $ m' $ run only over the $ l$ internal lines. The differential operators $\Delta_{\text{rr}}$ starting from an half-edge $  mA $ were defined in \eqref{rerouting} and depend on the energy variables in the diagram; it should be understood that these are defined on a common choice of energy flow out of the diagram as depicted in Figure \ref{routing}.  Moreover, the differential operator associated with this vertex at the final step is trivial, as all $\eta$ dependence has been accounted for, $\Delta_A = 1$, leaving only tensor and momentum contractions to be multiplied out front.  The final ingredient to be accounted for is the collapsing of edges.  The collapsing proceeds in the same way as described previously.  In all cases, the vertex in \eqref{replacementn3} implies $d_A = 4$, and the specific contractions depend on the vertices to which $A$ is connected. Below we apply this prescription to four-point exchange diagrams.  As an example, consider the following time integral:
\begin{equation}\label{6p7}
	I = \int d\eta_1 d\eta_2 e^{iq_1\eta_1}e^{iq_2\eta_2}\left[\eta_1^{-1}\partial_{\eta_1}-\frac{k_1k_2}{p}  \right]\left[\eta_2^{-1}\partial_{\eta_2}-\frac{k_3k_4}{p} \right]G_{\phi\phi}(\eta_1, \eta_2, p)  
\end{equation}
This is the time integral relevant for graviton exchange in minimal coupling to gravity.  We have applied the integration by part discussed above to both vertices.  Note that the term with two time derivatives $\partial_{\eta_1}\partial_{\eta_2}$ and the term without any time derivative will correspond to $\phi'\phi'$ and $\phi\phi$ propagators respectively.  These propagators display delta functions that need to be accounted for by the procedure of collapsing. In the next section we will use the differential operator prescription to compute the wavefunction for this process.

\subsection{Example: Scalars exchanging a graviton}

A massless scalar $\phi$ minimally coupled to gravity has the following interaction vertex with transverse-traceless tensor fluctuations $ \gamma_{ij}$:
\begin{equation}
	S_{\text{int}} = \int\frac{d\eta}{\eta^2}  d^3 x \gamma^{ij}\partial_i \phi \partial_j \phi\,.
\end{equation}
\begin{figure}
	\begin{center}
		\begin{tikzpicture}
			\coordinate (v1) at (-3, 0);
			\coordinate (v2) at (3, 0);
			\coordinate (b4) at (4, 2);
			\coordinate (b3) at (2, 2);
			\coordinate (b2) at (-2, 2);
			\coordinate (b1) at (-4, 2);
			\coordinate (vp) at (0, -.5);
			\draw[thick] (b1) -- (v1) node[anchor=east]{$  q_{1} $} -- (b2);
			\draw[thick] (b3)  --  (v2) node[anchor=west]{$  q_{2} $} -- (b4);
			\draw[thick,decorate, decoration={snake, segment length=3.15mm, pre length=0mm, post length=0mm}, domain=205:335] plot ({3.31*cos(\x)},{.64+1.5*sin(\x)});
			\draw[very thick] (-6, 2) -- (6, 2);
			\node at (0, -1.5) {$\gamma_{ij}(\v{p})$};
			\node at (-4, 2.5) {$\phi(\v{k}_{1})$};
			\node at (-2, 2.5) {$\phi(\v{k}_{2})$};
			\node at (2, 2.5) {$\phi(\v{k}_{3})$};
			\node at (4, 2.5) {$\phi(\v{k}_{4})$};
		\end{tikzpicture}
	\end{center}
	\caption{The $ s$-channel graviton-exchange contribution to $ \psi_{4}$. \label{ffgff}}
	\label{scalargraviton}
\end{figure}
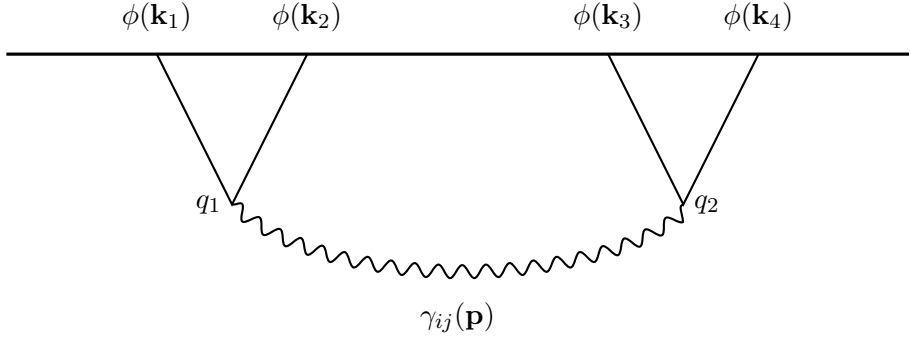
In the following we compute the contribution to the four-scalar wavefunction coefficient $ \psi_{4}$ from the $ s$-channel exchange of a graviton and two insertions of this interaction, as depicted in Figure \ref{ffgff}. This was previously computed in \cite{Seery:2008ax} using the bulk representation and in \cite{Ghosh:2014kba} for the wavefunction (see also \cite{COT}). Let's use the above prescription for one of the two interaction vertices. We notice that each vertex has a single internal line attached, so we should choose $ l=1 $ in \eqref{deltagrav}. This implies that both terms in \eqref{deltagrav} are non-vanishing. The left vertex with $  q_{1}=k_{1}+k_{2} $ energy gives the differential operator
\begin{align}
	\Delta_{\text{grav}}^{\text{Left}}= p-\frac{e_{2}(k_{1},k_{2})}{p}(2+q_1\partial_{q_1})=p-\frac{k_1k_2}{p}(2+q_1\partial_{q_1})\,,
\end{align} 
where we have use the vertex-side definition of $ \Drr$ in \eqref{rerouting}. Combining the differential operators of each vertex one finds
\begin{align}
	\Delta_\text{grav} &=	\Delta_{\text{grav}}^{\text{Left}} \,	\Delta_{\text{grav}}^{\text{Right}} \\
	&=  \left[p-\frac{k_1k_2}{p}(2+q_1\partial_{q_1}) \right]\left[p-\frac{k_3k_4}{p}(2+q_2\partial_{q_2}) \right]\,,
\end{align}
where we cut our momentum flow along the single internal line. This differential operator fully accounts for the exchanged time integral.  We finally account for collapsed contributions, of which there are two: one of the $\braket{\phi'\phi'}$ type, corresponding to the term with two time derivative in \eqref{6p7}, and one of the $\braket{\phi\phi}$ type, corresponding to the term without any time derivatives.  These contribute in the same way as discussed in Section \ref{sec:4}, with two time derivatives needed on the $\braket{\phi\phi}$ collapse.  Therefore we have 
\begin{align}
\text{collapsed contribution: } -\left( 1+ \frac{k_1k_2k_3k_4}{p^{2}}\partial_{q_1}\partial_{q_{2}} \right)\psiflat_{1}\,,
\end{align}
where $  \psiflat_{1}=k_{T}^{-1} $ with $  k_{T}=\sum k_{a}=q_{1}+q_{2} $. Putting together all contributions we find
\begin{align}
	\psi_{4\phi} &=F \left[-(1+ k_1k_2k_3k_4 \partial_{q_1}\partial_{q_{2}})\psiflat_{1} +\Delta_{\text{grav}}^{\text{Left}} \Delta_{\text{grav}}^{\text{Right}}  \psiflat_{2}\right]\\
	&=F\left[-\frac{1}{k_T}-\frac{2k_1 k_2 k_3 k_4}{k_T^3 p_1^2} +\left(p_1-\frac{k_1k_2}{p_1}(2+q_1\partial_{q_1}) \right)\left(p_1-\frac{k_3k_4}{p_1}(2+q_2\partial_{q_2}) \right)\psiflat\right]\,,
\end{align}
where $\psiflat$ is the flatspace seed function in \eqref{twosite}, and the vertex is given by
\begin{align}
F= \sum_{s=\pm} \epsilon_{ij}^{s}(\v{p})\epsilon^{s}_{kl}(-\v{p})k_1^ik_2^jk_3^kk_4^l\,,
\end{align} 
with $ \v{p}$ he internal momentum $ \v{p}=\v{k}_{1}+\v{k}_{2}$. 
This sum can be re-written without any reference to the polarization tensors $ \epsilon_{ij}^{s}(\v{p})$ using (see e.g. \cite{Goon:2018fyu})
\begin{align}\label{again}
\sum_{s=\pm2}&=\epsilon_{ij}^{s}(\v{p})\epsilon^{s}_{kl}(-\v{p})= \pi_{ik}\pi_{jl}+\pi_{il}\pi_{jk}-\pi_{ij}\pi_{kl}\,, &
\pi_{ij}&\equiv\delta_{ij}-\frac{p_{i}p_{j}}{p^{2}}\,,
\end{align}
This gives
\begin{align}
\epsilon_{ij}^{s}\epsilon^{s}_{kl}k_1^ik_2^jk_3^kk_4^l&=\left(  \v{k}_{1}\cdot \v{k}_{3}+\frac{\v{k}_{1}\cdot \v{p} \,\v{k}_{3}\cdot \v{p}}{p^{2}}\right)\left(  \v{k}_{2}\cdot \v{k}_{4}+\frac{\v{k}_{2}\cdot \v{p}\,\v{k}_{4}\cdot \v{p}}{p^{2}}\right)\\
&+\left(  \v{k}_{1}\cdot \v{k}_{4}+\frac{\v{k}_{1}\cdot \v{p}\,\v{k}_{4}\cdot \v{p}}{p^{2}}\right)\left(  \v{k}_{2}\cdot \v{k}_{3}+\frac{\v{k}_{2}\cdot \v{p}\,\v{k}_{3}\cdot \v{p}}{p^{2}}\right)\\
&-\left(  \v{k}_{1}\cdot \v{k}_{2}+\frac{\v{k}_{1}\cdot \v{p}\,\v{k}_{2}\cdot \v{p}}{p^{2}}\right)\left(  \v{k}_{3}\cdot \v{k}_{4}+\frac{\v{k}_{3}\cdot \v{p}\,\v{k}_{4}\cdot \v{p}}{p^{2}}\right)\,.
\end{align}


\subsection{Example: Gravitons exchanging a graviton}

If we consider the tensor self-interaction at third order we have the vertex
\begin{equation}
	\int \frac{d\eta }{\eta^2} d^3 x \left(\gamma_{ik} \gamma_{jl}-\frac{1}{2} \gamma_{ij}\gamma_{kl}\right)\partial_k \partial_l \gamma_{ij}\,, 
\end{equation}
which produces an identical time integral as the one in the previous example but a different polarization structure out front. In particular one has
\begin{equation}\label{notonly}
	\psi_{4\gamma} = F\left[-\frac{1}{k_T}-\frac{2k_1 k_2 k_3 k_4}{k_T^3 p_1^2} +\left(p_1-\frac{k_1k_2}{p_1}(2+q_1\partial_{q_1}) \right)\left(p_1-\frac{k_3k_4}{p_1}(2+q_2\partial_{q_2}) \right)\psiflat_{2}\right]\,,
\end{equation}
where the vertex now is 
\begin{align}
F=\epsilon_{ii'}^{s_{1}}(\v{k}_{1})\epsilon_{jj'}^{s_{2}}(\v{k}_{2})t_{ijl}t_{i'j'l'}\left[ \sum_{s=\pm} \epsilon_{ll'}^{s}(\v{p})\epsilon^{s}_{kk'}(-\v{p}) \right]\epsilon_{mm'}^{s_{3}}(\v{k}_{3})\epsilon_{nn'}^{s_{4}}(\v{k}_{4}) \tilde t_{kmn} \tilde t_{k'm'n'}\,,
\end{align}
with
\begin{align}
t_{ijl}&=\delta_{lj}k^{i}_{2}+\delta_{il}p^{j}+\delta_{ij}k^{l}_{1}\,, & \tilde t_{ijl}&=\delta_{lj}k^{i}_{3}+\delta_{il}k^{j}_{4}+\delta_{ij}p^{l}\,.
\end{align}
If desired, the sum over polarization tensors can again be re-written as in \eqref{again}. Notice that \eqref{notonly} is \textit{not} the only contributions to the graviton quartic wavefunction coefficient and in fact it is not even gauge invariant per se. In the gauge we are working with, namely that of \cite{Maldacena:2002vr}, the missing terms are quartic contact contributions resulting from $  R^{(3)} $, the square of the extrinsic curvature and the integrating out of the lapse and the shift at second order. We leave a full computation of the missing terms for future work. The application of equation \ref{replacement} and subsequent expansion in terms of differential operators acting on seed functions can be carried out for arbitrary trees.

 
\section{Conclusions and outlook}\label{sec:end}

In this work, we have provided a differential representation of tree-level wavefunction coefficients for massless scalars and gravitons in de Sitter. Our representation is valid for general boost breaking interactions satisfying the bound in \eqref{bound}. In Section \ref{sec:minimal} we have shown how to generalise this representation to cover also gravitational interactions in GR with minimal coupling, which violate \eqref{bound}. The differential representation involves purely algebraic operations and so side-steps the need to calculate the nested time integrals that arise in the usual bulk representation based on Feyman-Witten rules. We have provided a series of concrete examples to show how the differential representation is computationally less laborious than the time-integral representation, especially for diagrams with two or more internal propagators. Also, the final result in terms of differential operators takes a relative compact form. On the one hand, this form can already be used for algebraic manipulations. On the other hand, it can be expanded into a fully explicit rational function in a fraction of a second using computer software such as Mathematica. Finally, the differential representation has the property to completely eliminate time from the calculation, which might make it more attune to a potential holographic description of perturbative QFT in de Sitter.

This research can be extended as follows:
\begin{itemize}
\item It would be interesting to understand how bulk locality constrains the form of the differential operator that acts on the flat spacetime seed $  \psiflat $ to give the dS wavefunction. This might have some relation to the recently discussed manifestly local test \cite{MLT}.
\item One should be able to build a minor modification of our construction that computes the wavefunction for boost-breaking interactions of a conformally coupled scalar. Related to this, it would be nice to understand the connection of our procedure to the formalism of the cosmological polytopes of \cite{cosmopoly,Arkani-Hamed:2018bjr,Benincasa:2019vqr}. 
\end{itemize}

 
\section*{Acknowledgements}

We would like to thank Harry Goodhew, Nima Arkani-Hamed, Sebastian Mizera, and Guilherme Pimentel for useful discussions and S. Jazayeri, M.H.G. Lee, S. Melville and J. Supe\l{} for comments on the draft. E.P. has been supported in part
by the research program VIDI with Project No. 680-47-535, which is (partly) financed by the Netherlands Organisation for Scientific Research (NWO). This work has been partially supported by STFC consolidated grant ST/T000694/1.

 \appendix
 
 
 \section{Derivation of Propagator Relations}\label{app:A}
 Here we derive the propagator relations upon which our analysis rests.  Consider the massless propagator in de Sitter and flat space respectively,
 \begin{align}
 	G(\eta, \eta', k) &= \frac{i}{2k^3}\left[\phi_k^\star(\eta)\phi_k(\eta')\theta(\eta-\eta')+\phi_k(\eta)\phi_k^\star(\eta')\theta(\eta'-\eta
 	)-\phi_k(\eta)\phi_k(\eta') \right]\,, \vspace{5mm}\\ \nonumber\\
 	G_{\text{flat}}(\eta, \eta', k) &= \frac{i}{2k}\left[e^{-ik(\eta -\eta')}\theta(\eta-\eta')+e^{-ik(\eta'-\eta)}\theta(\eta'-\eta)-e^{ik(\eta+\eta')} \right]\,,
 \end{align}
 where the de Sitter propagator is composed of the massless mode functions 
 \begin{equation}
 	\phi_k(\eta) = (1-ik)e^{ik\eta}\,.
 \end{equation}
 Now consider the two time-derivative propagator $  G_{\phi'\phi'} = \partial_\eta \partial_{\eta'}G$  integrated against some function $F(\eta, \eta')$
 \begin{equation}
 	\int\limits_{-\infty}^0d\eta d\eta' F(\eta, \eta')\partial_\eta \partial_{\eta'}G(\eta, \eta', k)\,.
 \end{equation}
When the time derivatives hit both mode functions one gets $k^2 \eta\eta'G_{\text{flat}}(\eta, \eta', k)$. Now we turn to the terms hitting the $\theta$-functions.  First we consider terms where a single $\theta$-function is hit.  This contributes
 \begin{equation}
 	\frac{2i\delta(\eta-\eta')}{2k}\left[\phi^\star_k(\eta)\partial_{\eta'}\phi_k(\eta')-\phi_k(\eta)\partial_{\eta'}\phi^\star_k(\eta') \right] = -2\eta \eta'\delta(\eta-\eta')\,.
 \end{equation}
 Finally, we consider when both $\theta$-functions are hit with the derivative, which gives
 \begin{equation}
 	\int d\eta d\eta' F(\eta, \eta') \frac{i}{2k^3}\left[\phi_k(\eta)\phi_k^\star(\eta')\partial_\eta \delta(\eta'-\eta)- \phi_k^\star(\eta)\phi_k(\eta')\partial_\eta \delta(\eta'-\eta)\right]\,.
\end{equation}
We now integrate by parts in $\eta$.  The non-vanishing contribution comes from the derivative hitting the mode functions.  This gives us 
\begin{align}
	\int d\eta d\eta' F(\eta, \eta') \frac{i}{2k^3}\left[\phi_k(\eta')\partial_\eta \phi_k^\star(\eta)-\phi_k^\star(\eta') \partial_\eta \phi_k(\eta) \right]\delta(\eta-\eta') = \int d\eta d\eta'F(\eta, \eta') [ \eta \eta'\delta(\eta-\eta')]\,.
\end{align}
Summing the contributions we recover \eqref{propspp},
\begin{align}
	\partial_\eta\partial_{\eta'}G(\eta, \eta', k) = \eta\eta'\left[k^2G_{\text{flat}}(\eta, \eta', k)-\delta(\eta-\eta') \right]\,.
\end{align}
Now we consider generating the $  \phi\phi $ and mixed $  \phi' \phi $ de Sitter propagators from the flat space propagator.  First note that the mixed propagator $  G_{\phi'\phi} $ is
\begin{align}
	\partial_\eta G(\eta, \eta') = \frac{i}{2k}\left[\eta(1-ik\eta')e^{-ik(\eta-\eta')}\theta(\eta-\eta')+\eta(1+ik\eta')e^{-ik(\eta'-\eta)}\theta(\eta'-\eta)-\eta(1-ik\eta')e^{ik(\eta+\eta')} \right]\,.
\end{align}
Notice that the single-derivative hitting the $\theta$-function vanishes on the support of the $\delta$-function it generates.  Given the mode function relation 
\begin{align}
	(1-\eta\partial_\eta)e^{ik\eta} = (1-ik\eta)e^{ik\eta}\,,
\end{align}
we have 
\begin{align}
	G_{\phi'\phi}&=\partial_\eta G(\eta, \eta', k) = (1-\eta'\partial_{\eta'})G_{\text{flat}}(\eta, \eta', k)\,, \\ \nonumber \\
	 G(\eta, \eta')&= \frac{1}{k^2} \left[(1-\eta\partial_\eta)(1-\eta'\partial_{\eta'})G_{\text{flat}}(\eta, \eta', k)+\eta\eta'\delta(\eta-\eta') \right]\,,\label{dsprop}
\end{align}
where the $\delta$-function in the above expression for $  G $ comes from analogous integration by parts manipulations on the term with two derivatives.

\bibliographystyle{JHEP}
\bibliography{refs}

\providecommand{\href}[2]{#2}\begingroup\raggedright\begin{thebibliography}{10}

\bibitem{Positivity1}
A.~Adams, N.~Arkani-Hamed, S.~Dubovsky, A.~Nicolis and R.~Rattazzi,
  \emph{{Causality, analyticity and an IR obstruction to UV completion}},
  \href{https://doi.org/10.1088/1126-6708/2006/10/014}{\emph{JHEP} {\bfseries
  10} (2006) 014} [\href{https://arxiv.org/abs/hep-th/0602178}{{\ttfamily
  hep-th/0602178}}].

\bibitem{Paulos:2016fap}
M.F.~Paulos, J.~Penedones, J.~Toledo, B.C.~van Rees and P.~Vieira, \emph{{The
  S-matrix bootstrap. Part I: QFT in AdS}},
  \href{https://doi.org/10.1007/JHEP11(2017)133}{\emph{JHEP} {\bfseries 11}
  (2017) 133} [\href{https://arxiv.org/abs/1607.06109}{{\ttfamily
  1607.06109}}].

\bibitem{cosmopoly}
N.~Arkani-Hamed, P.~Benincasa and A.~Postnikov, \emph{{Cosmological Polytopes
  and the Wavefunction of the Universe}},
  \href{https://arxiv.org/abs/1709.02813}{{\ttfamily 1709.02813}}.

\bibitem{Goon:2018fyu}
G.~Goon, K.~Hinterbichler, A.~Joyce and M.~Trodden, \emph{{Shapes of gravity:
  Tensor non-Gaussianity and massive spin-2 fields}},
  \href{https://doi.org/10.1007/JHEP10(2019)182}{\emph{JHEP} {\bfseries 10}
  (2019) 182} [\href{https://arxiv.org/abs/1812.07571}{{\ttfamily
  1812.07571}}].

\bibitem{Arkani-Hamed:2018bjr}
N.~Arkani-Hamed and P.~Benincasa, \emph{{On the Emergence of Lorentz Invariance
  and Unitarity from the Scattering Facet of Cosmological Polytopes}},
  \href{https://arxiv.org/abs/1811.01125}{{\ttfamily 1811.01125}}.

\bibitem{Benincasa:2018ssx}
P.~Benincasa, \emph{{From the flat-space S-matrix to the Wavefunction of the
  Universe}},  \href{https://arxiv.org/abs/1811.02515}{{\ttfamily 1811.02515}}.

\bibitem{CosmoBootstrap1}
N.~Arkani-Hamed, D.~Baumann, H.~Lee and G.L.~Pimentel, \emph{{The Cosmological
  Bootstrap: Inflationary Correlators from Symmetries and Singularities}},
  \href{https://doi.org/10.1007/JHEP04(2020)105}{\emph{JHEP} {\bfseries 04}
  (2020) 105} [\href{https://arxiv.org/abs/1811.00024}{{\ttfamily
  1811.00024}}].

\bibitem{Baumann:2019oyu}
D.~Baumann, C.~Duaso~Pueyo, A.~Joyce, H.~Lee and G.L.~Pimentel, \emph{{The
  Cosmological Bootstrap: Weight-Shifting Operators and Scalar Seeds}},
  \href{https://arxiv.org/abs/1910.14051}{{\ttfamily 1910.14051}}.

\bibitem{Sleight:2019mgd}
C.~Sleight, \emph{{A Mellin Space Approach to Cosmological Correlators}},
  \href{https://doi.org/10.1007/JHEP01(2020)090}{\emph{JHEP} {\bfseries 01}
  (2020) 090} [\href{https://arxiv.org/abs/1906.12302}{{\ttfamily
  1906.12302}}].

\bibitem{Sleight:2019hfp}
C.~Sleight and M.~Taronna, \emph{{Bootstrapping Inflationary Correlators in
  Mellin Space}}, \href{https://doi.org/10.1007/JHEP02(2020)098}{\emph{JHEP}
  {\bfseries 02} (2020) 098}
  [\href{https://arxiv.org/abs/1907.01143}{{\ttfamily 1907.01143}}].

\bibitem{Benincasa:2019vqr}
P.~Benincasa, \emph{{Cosmological Polytopes and the Wavefuncton of the Universe
  for Light States}},  \href{https://arxiv.org/abs/1909.02517}{{\ttfamily
  1909.02517}}.

\bibitem{Bzowski:2019kwd}
A.~Bzowski, P.~McFadden and K.~Skenderis, \emph{{Conformal $n$-point functions
  in momentum space}},
  \href{https://doi.org/10.1103/PhysRevLett.124.131602}{\emph{Phys. Rev. Lett.}
  {\bfseries 124} (2020) 131602}
  [\href{https://arxiv.org/abs/1910.10162}{{\ttfamily 1910.10162}}].

\bibitem{Baumann:2020dch}
D.~Baumann, C.~Duaso~Pueyo, A.~Joyce, H.~Lee and G.L.~Pimentel, \emph{{The
  Cosmological Bootstrap: Spinning Correlators from Symmetries and
  Factorization}},  \href{https://arxiv.org/abs/2005.04234}{{\ttfamily
  2005.04234}}.

\bibitem{Green:2020ebl}
D.~Green and E.~Pajer, \emph{{On the Symmetries of Cosmological
  Perturbations}},  \href{https://arxiv.org/abs/2004.09587}{{\ttfamily
  2004.09587}}.

\bibitem{COT}
H.~Goodhew, S.~Jazayeri and E.~Pajer, \emph{{The Cosmological Optical
  Theorem}},  \href{https://arxiv.org/abs/2009.02898}{{\ttfamily 2009.02898}}.

\bibitem{Sleight:2020obc}
C.~Sleight and M.~Taronna, \emph{{From AdS to dS Exchanges: Spectral
  Representation, Mellin Amplitudes and Crossing}},
  \href{https://arxiv.org/abs/2007.09993}{{\ttfamily 2007.09993}}.

\bibitem{BBBB}
E.~Pajer, \emph{{Building a Boostless Bootstrap for the Bispectrum}},
  \href{https://doi.org/10.1088/1475-7516/2021/01/023}{\emph{JCAP} {\bfseries
  01} (2021) 023} [\href{https://arxiv.org/abs/2010.12818}{{\ttfamily
  2010.12818}}].

\bibitem{MLT}
S.~Jazayeri, E.~Pajer and D.~Stefanyszyn, \emph{{From Locality and Unitarity to
  Cosmological Correlators}},
  \href{https://arxiv.org/abs/2103.08649}{{\ttfamily 2103.08649}}.

\bibitem{Isono:2020qew}
H.~Isono, H.M.~Liu and T.~Noumi, \emph{{Wavefunctions in dS/CFT revisited:
  principal series and double-trace deformations}},
  \href{https://arxiv.org/abs/2011.09479}{{\ttfamily 2011.09479}}.

\bibitem{Melville:2021lst}
S.~Melville and E.~Pajer, \emph{{Cosmological Cutting Rules}},
  \href{https://doi.org/10.1007/JHEP05(2021)249}{\emph{JHEP} {\bfseries 05}
  (2021) 249} [\href{https://arxiv.org/abs/2103.09832}{{\ttfamily
  2103.09832}}].

\bibitem{Goodhew:2021oqg}
H.~Goodhew, S.~Jazayeri, M.H.~Gordon~Lee and E.~Pajer, \emph{{Cutting
  Cosmological Correlators}},
  \href{https://arxiv.org/abs/2104.06587}{{\ttfamily 2104.06587}}.

\bibitem{Bonifacio:2021azc}
J.~Bonifacio, E.~Pajer and D.-G.~Wang, \emph{{From Amplitudes to Contact
  Cosmological Correlators}},
  \href{https://arxiv.org/abs/2106.15468}{{\ttfamily 2106.15468}}.

\bibitem{DiPietro:2021sjt}
L.~Di~Pietro, V.~Gorbenko and S.~Komatsu, \emph{{Analyticity and Unitarity for
  Cosmological Correlators}},
  \href{https://arxiv.org/abs/2108.01695}{{\ttfamily 2108.01695}}.

\bibitem{Sleight:2021plv}
C.~Sleight and M.~Taronna, \emph{{From dS to AdS and back}},
  \href{https://arxiv.org/abs/2109.02725}{{\ttfamily 2109.02725}}.

\bibitem{Hogervorst:2021uvp}
M.~Hogervorst, J.a.~Penedones and K.S.~Vaziri, \emph{{Towards the
  non-perturbative cosmological bootstrap}},
  \href{https://arxiv.org/abs/2107.13871}{{\ttfamily 2107.13871}}.

\bibitem{Meltzer:2021zin}
D.~Meltzer, \emph{{The Inflationary Wavefunction from Analyticity and
  Factorization}},  \href{https://arxiv.org/abs/2107.10266}{{\ttfamily
  2107.10266}}.

\bibitem{Sleight:2021iix}
C.~Sleight and M.~Taronna, \emph{{On the consistency of (partially-)massless
  matter couplings in de Sitter space}},
  \href{https://arxiv.org/abs/2106.00366}{{\ttfamily 2106.00366}}.

\bibitem{Cespedes:2020xqq}
S.~C\'espedes, A.-C.~Davis and S.~Melville, \emph{{On the time evolution of
  cosmological correlators}},
  \href{https://arxiv.org/abs/2009.07874}{{\ttfamily 2009.07874}}.

\bibitem{Gomez:2021qfd}
H.~Gomez, R.L.~Jusinskas and A.~Lipstein, \emph{{Cosmological Scattering
  Equations}},  \href{https://arxiv.org/abs/2106.11903}{{\ttfamily
  2106.11903}}.

\bibitem{Cabass:2021fnw}
G.~Cabass, E.~Pajer, D.~Stefanyszyn and J.~Supe\l{}, \emph{{Bootstrapping Large
  Graviton non-Gaussianities}},
  \href{https://arxiv.org/abs/2109.10189}{{\ttfamily 2109.10189}}.

\bibitem{Maldacena:2002vr}
J.M.~Maldacena, \emph{{Non-Gaussian features of primordial fluctuations in
  single field inflationary models}},
  \href{https://doi.org/10.1088/1126-6708/2003/05/013}{\emph{JHEP} {\bfseries
  05} (2003) 013} [\href{https://arxiv.org/abs/astro-ph/0210603}{{\ttfamily
  astro-ph/0210603}}].

\bibitem{WFCtoCorrelators1}
D.~Anninos, T.~Anous, D.Z.~Freedman and G.~Konstantinidis, \emph{{Late-time
  Structure of the Bunch-Davies De Sitter Wavefunction}},
  \href{https://doi.org/10.1088/1475-7516/2015/11/048}{\emph{JCAP} {\bfseries
  11} (2015) 048} [\href{https://arxiv.org/abs/1406.5490}{{\ttfamily
  1406.5490}}].

\bibitem{Mata:2012bx}
I.~Mata, S.~Raju and S.~Trivedi, \emph{{CMB from CFT}},
  \href{https://doi.org/10.1007/JHEP07(2013)015}{\emph{JHEP} {\bfseries 07}
  (2013) 015} [\href{https://arxiv.org/abs/1211.5482}{{\ttfamily 1211.5482}}].

\bibitem{Bzowski:2013sza}
A.~Bzowski, P.~McFadden and K.~Skenderis, \emph{{Implications of conformal
  invariance in momentum space}},
  \href{https://doi.org/10.1007/JHEP03(2014)111}{\emph{JHEP} {\bfseries 03}
  (2014) 111} [\href{https://arxiv.org/abs/1304.7760}{{\ttfamily 1304.7760}}].

\bibitem{Kundu:2014gxa}
N.~Kundu, A.~Shukla and S.P.~Trivedi, \emph{{Constraints from Conformal
  Symmetry on the Three Point Scalar Correlator in Inflation}},
  \href{https://doi.org/10.1007/JHEP04(2015)061}{\emph{JHEP} {\bfseries 04}
  (2015) 061} [\href{https://arxiv.org/abs/1410.2606}{{\ttfamily 1410.2606}}].

\bibitem{Kundu:2015xta}
N.~Kundu, A.~Shukla and S.P.~Trivedi, \emph{{Ward Identities for Scale and
  Special Conformal Transformations in Inflation}},
  \href{https://doi.org/10.1007/JHEP01(2016)046}{\emph{JHEP} {\bfseries 01}
  (2016) 046} [\href{https://arxiv.org/abs/1507.06017}{{\ttfamily
  1507.06017}}].

\bibitem{Conjecture}
D.~Baumann, D.~Green, H.~Lee and R.A.~Porto, \emph{{Signs of Analyticity in
  Single-Field Inflation}},
  \href{https://doi.org/10.1103/PhysRevD.93.023523}{\emph{Phys. Rev.}
  {\bfseries D93} (2016) 023523}
  [\href{https://arxiv.org/abs/1502.07304}{{\ttfamily 1502.07304}}].

\bibitem{TanguyScott}
T.~Grall and S.~Melville, \emph{{Inflation in Motion: Unitarity Constraints in
  Effective Field Theories with Broken Lorentz Symmetry}},
  \href{https://arxiv.org/abs/2005.02366}{{\ttfamily 2005.02366}}.

\bibitem{PSS}
E.~Pajer, D.~Stefanyszyn and J.~Supe\l{}, \emph{{The Boostless Bootstrap:
  Amplitudes without Lorentz boosts}},
  \href{https://doi.org/10.1007/JHEP12(2020)198}{\emph{JHEP} {\bfseries 12}
  (2020) 198} [\href{https://arxiv.org/abs/2007.00027}{{\ttfamily
  2007.00027}}].

\bibitem{DSJS}
D.~Stefanyszyn and J.~Supe\l{}, \emph{{The Boostless Bootstrap and BCFW
  Momentum Shifts}},  \href{https://arxiv.org/abs/2009.14289}{{\ttfamily
  2009.14289}}.

\bibitem{Positivity2}
T.~Grall and S.~Melville, \emph{{Positivity Bounds without Boosts}},
  \href{https://arxiv.org/abs/2102.05683}{{\ttfamily 2102.05683}}.

\bibitem{Melville:2019wyy}
S.~Melville and J.~Noller, \emph{{Positivity in the Sky: Constraining dark
  energy and modified gravity from the UV}},
  \href{https://doi.org/10.1103/PhysRevD.101.021502}{\emph{Phys. Rev. D}
  {\bfseries 101} (2020) 021502}
  [\href{https://arxiv.org/abs/1904.05874}{{\ttfamily 1904.05874}}].

\bibitem{Stefanyszyn:2020kay}
D.~Stefanyszyn and J.~Supe\l{}, \emph{{The Boostless Bootstrap and BCFW
  Momentum Shifts}}, \href{https://doi.org/10.1007/JHEP03(2021)091}{\emph{JHEP}
  {\bfseries 03} (2021) 091}
  [\href{https://arxiv.org/abs/2009.14289}{{\ttfamily 2009.14289}}].

\bibitem{Celoria:2021vjw}
M.~Celoria, P.~Creminelli, G.~Tambalo and V.~Yingcharoenrat, \emph{{Beyond
  perturbation theory in inflation}},
  \href{https://doi.org/10.1088/1475-7516/2021/06/051}{\emph{JCAP} {\bfseries
  06} (2021) 051} [\href{https://arxiv.org/abs/2103.09244}{{\ttfamily
  2103.09244}}].

\bibitem{Creminelli:2006xe}
P.~Creminelli, M.A.~Luty, A.~Nicolis and L.~Senatore, \emph{{Starting the
  Universe: Stable Violation of the Null Energy Condition and Non-standard
  Cosmologies}},
  \href{https://doi.org/10.1088/1126-6708/2006/12/080}{\emph{JHEP} {\bfseries
  12} (2006) 080} [\href{https://arxiv.org/abs/hep-th/0606090}{{\ttfamily
  hep-th/0606090}}].

\bibitem{Cheung:2007st}
C.~Cheung, P.~Creminelli, A.L.~Fitzpatrick, J.~Kaplan and L.~Senatore,
  \emph{{The Effective Field Theory of Inflation}},
  \href{https://doi.org/10.1088/1126-6708/2008/03/014}{\emph{JHEP} {\bfseries
  03} (2008) 014} [\href{https://arxiv.org/abs/0709.0293}{{\ttfamily
  0709.0293}}].

\bibitem{Bordin:2018pca}
L.~Bordin, P.~Creminelli, A.~Khmelnitsky and L.~Senatore, \emph{{Light
  Particles with Spin in Inflation}},
  \href{https://doi.org/10.1088/1475-7516/2018/10/013}{\emph{JCAP} {\bfseries
  10} (2018) 013} [\href{https://arxiv.org/abs/1806.10587}{{\ttfamily
  1806.10587}}].

\bibitem{SolidInflation}
S.~Endlich, A.~Nicolis and J.~Wang, \emph{{Solid Inflation}},
  \href{https://doi.org/10.1088/1475-7516/2013/10/011}{\emph{JCAP} {\bfseries
  10} (2013) 011} [\href{https://arxiv.org/abs/1210.0569}{{\ttfamily
  1210.0569}}].

\bibitem{Baumann:2021fxj}
D.~Baumann, W.-M.~Chen, C.~Duaso~Pueyo, A.~Joyce, H.~Lee and G.L.~Pimentel,
  \emph{{Linking the Singularities of Cosmological Correlators}},
  \href{https://arxiv.org/abs/2106.05294}{{\ttfamily 2106.05294}}.

\bibitem{Senatore:2009gt}
L.~Senatore, K.M.~Smith and M.~Zaldarriaga, \emph{{Non-Gaussianities in Single
  Field Inflation and their Optimal Limits from the WMAP 5-year Data}},
  \href{https://doi.org/10.1088/1475-7516/2010/01/028}{\emph{JCAP} {\bfseries
  01} (2010) 028} [\href{https://arxiv.org/abs/0905.3746}{{\ttfamily
  0905.3746}}].

\bibitem{Henningson:1998gx}
M.~Henningson and K.~Skenderis, \emph{{The Holographic Weyl anomaly}},
  \href{https://doi.org/10.1088/1126-6708/1998/07/023}{\emph{JHEP} {\bfseries
  07} (1998) 023} [\href{https://arxiv.org/abs/hep-th/9806087}{{\ttfamily
  hep-th/9806087}}].

\bibitem{Seery:2008ax}
D.~Seery, M.S.~Sloth and F.~Vernizzi, \emph{{Inflationary trispectrum from
  graviton exchange}},
  \href{https://doi.org/10.1088/1475-7516/2009/03/018}{\emph{JCAP} {\bfseries
  03} (2009) 018} [\href{https://arxiv.org/abs/0811.3934}{{\ttfamily
  0811.3934}}].

\bibitem{Ghosh:2014kba}
A.~Ghosh, N.~Kundu, S.~Raju and S.P.~Trivedi, \emph{{Conformal Invariance and
  the Four Point Scalar Correlator in Slow-Roll Inflation}},
  \href{https://doi.org/10.1007/JHEP07(2014)011}{\emph{JHEP} {\bfseries 07}
  (2014) 011} [\href{https://arxiv.org/abs/1401.1426}{{\ttfamily 1401.1426}}].

\end{thebibliography}\endgroup

\end{document}